\documentclass[prb,reprint,amsmath,amssymb,floatfix,superscriptaddress]{revtex4-1}
\usepackage{graphicx}
\usepackage{hyperref}
\usepackage{enumerate}
\renewcommand{\vec}[1]{\mathbf{#1}}

\begin{document}

\title{Intrinsic ac anomalous Hall effect of nonsymmorphic chiral superconductors with an application to $\mathrm{UPt_3}$}

\author{Zhiqiang Wang}
\affiliation{Department of Physics and Astronomy, McMaster University, Hamilton, Ontario, L8S 4M1, Canada}
\author{John Berlinsky}
\affiliation{Department of Physics and Astronomy, McMaster University, Hamilton, Ontario, L8S 4M1, Canada}
\author{Gertrud Zwicknagl}
\affiliation{Institut f\"{u}r Mathematische Physik, Technische Universit\"{a}t Braunschweig, 38106 Braunschweig, Germany}
\author{Catherine Kallin}
\affiliation{Department of Physics and Astronomy, McMaster University, Hamilton, Ontario, L8S 4M1, Canada}
\affiliation{Canadian Institute for Advanced Research, Toronto, Ontario M5G 1Z8, Canada}
\date{\today}

\begin{abstract}
We identify an intrinsic mechanism of the anomalous Hall effect for non-symmorphic chiral superconductors.
This mechanism relies on  both a nontrivial multi-band chiral superconducting order parameter, which is a mixture of pairings of even and odd angular momentum channels, and a complex normal state inter-sublattice hopping, both of which are consequences
of the nonsymmorphic group symmetry of the underlying lattice. We apply this mechanism to the putative chiral superconducting phase of the heavy-fermion superconductor $\mathrm{UPt_3}$ and calculate
the anomalous ac Hall conductivity in a simplified two-band model.  From the ac Hall conductivity and optical data we estimate the polar Kerr rotation angle and compare it to the
measured results for $\mathrm{UPt_3}$ [E. R. Schemm \textit{et al.}, Science \textbf{345},190(2014)].
\end{abstract}


\maketitle
\section{Introduction} \label{sec:intro}
Understanding unconventional superconductors has been one of the central goals in condensed matter research.  Among the various unconventional superconductors,
chiral superconductors have attracted a great deal of attention in recent years, in part because they provide a platform to study the interplay between spontaneous symmetry
breaking and topology~\cite{Kallin2016}.  In a chiral superconductor, a Cooper pair
carries a nonzero relative orbital angular momentum whose projection along a certain direction is also nonzero. Choosing this direction as the angular momentum quantization
axis $z$, different chiral superconductors that are eigenstates of angular momentum can be characterized by the Cooper pair orbital angular momentum quantum numbers, $L=1,2,3,\dots$ and $L_z=\pm1,\pm2,\dots$.  A general chiral superconducting order, however, need not be an angular momentum eigenstate. For example, chiral f-wave may mix with chiral p-wave, etc.

One of the defining properties of a chiral superconductor is its spontaneous breaking of parity and time-reversal symmetry. As a consequence, there can be a nonzero anomalous
Hall effect (i.e., a Hall effect in the absence of an external magnetic field), which can be detected by polar Kerr effect measurements~\cite{Kapitulnik2009}.
Experimentally, a frequency dependent rotation angle between the polarization of incident and reflected light is measured. This Kerr angle, $\theta_K(\omega)$, is related to
the ac anomalous Hall conductivity, $\sigma_H(\omega)$, by~\cite{Argyres1955}
\begin{gather}
\theta_K (\omega)=\frac{4\pi }{\omega } \; \mathrm{Im} \bigg[  \frac{\sigma_H(\omega) }{n(n^2-1)}\bigg], \label{eq:Kerr}
\end{gather}
where $n$ is the frequency dependent index of refraction.  A nonzero Kerr signal has been observed in the superconducting phase of several unconventional
superconductors including $\mathrm{Sr_2 Ru O_4}$~\cite{Xia2006}, $\mathrm{UPt_3}$~\cite{Schemm2014}, $\mathrm{U Ru_2 Si_2}$~\cite{Schemm2015},
$\mathrm{Pr Os_4 Sb_{12}}$~\cite{Levenson-Falk2016}, and $\mathrm{Bi/Ni}$ bilayers~\cite{Gong2017}. $\mathrm{Sr_2 Ru O_4}$ is widely thought to be a chiral $p$-wave superconductor~\cite{Kallin2009,Mackenzie2017};
while the heavy fermion superconductor $\mathrm{UPt_3}$ is expected to be a chiral $f$-wave superconductor with $E_{2u}$ symmetry, corresponding to $L=3$, $L_z=\pm2$
in the continuum limit.~\cite{Sauls1994,Norman1992}

However, parity and time reversal symmetry breaking are necessary but not sufficient conditions for a nonzero anomalous Hall effect.
Breaking of additional symmetries, translation and particle-hole, are needed for a nonzero $\sigma_H(\omega)$.
Consequently, the size of the effect depends crucially on the mechanism by which these symmetries are broken.
As pointed out previously~\cite{Read2000,Roy2008,Lutchyn2009}, $\sigma_H(\omega)$ vanishes at all frequencies for a Galliean invariant chiral superconductor.
One way to break translation symmetry is by extrinsic impurity scattering, which has been studied by several groups in the context of $\mathrm{Sr_2 Ru O_4}$~\cite{Goryo2008,Lutchyn2009,Koenig2017}.
This impurity effect does not contribute to $\sigma_H$ in the lowest order Born approximation and therefore requires higher order scattering~\cite{Goryo2008}.
However, both $\mathrm{Sr_2 Ru O_4}$ and $\mathrm{UPt_3}$ are very clean, and it is not clear if the observed effect is due to disorder. Even without impurities, translation symmetry can be broken by certain intrinsic mechanisms, which turn out to be rather subtle. There have been two intrinsic mechanisms
proposed previously. One is based on a collective mode~\cite{Yip1992}, combined with the small but finite momentum of the incident photon and the breaking of inversion symmetry along the incident
external electro-magnetic wave propagation direction. However,
the estimated angle for this mechanism is too small to account for experiments~\cite{Xia2006}.
The other intrinsic mechanism invokes a multiband effect~\cite{Taylor2012,Taylor2013,Wysokinski2012,Gradhand2013,Mineev2014}, arising from structure within the crystal unit cell,  which also involves interband
pairing.  Here, we will study a generalization of this multi-band mechanism.

All of these theories (impurity effects, collective modes, and the multiband effect) have so far only been studied for the case of chiral $p$-wave superconductors.
This has led to a better understanding of the Kerr effect in $\mathrm{Sr_2 Ru O_4}$. However, $\mathrm{UPt_3}$ is thought to be a chiral $f$-wave superconductor
in its lower superconducting transition temperature phase. One might think that the conclusions obtained for the Kerr effect in a chiral $p$-wave superconductor can
be directly generalized to higher chirality superconductors with $|L_z|\ge 2$ without much difficulty. However, such a naive generalization
is problematic. As recent studies on non-topologically protected quantities, such as the integrated edge current~\cite{Huang2015} and the total orbital
angular momentum~\cite{Tada2015,Volovik2015},  have demonstrated explicitly, chiral superconductors with
$|L_z|\ge 2$ can behave very differently from
the chiral $p$-wave case. Given that the anomalous Hall conductivity $\sigma_H(\omega)$ is also a non-topologically protected quantity~\cite{Read2000,Lutchyn2009}, unlike
its thermal Hall counterpart, we expect that $\sigma_H(\omega)$ of
chiral superconductors with $|L_z| \ge 2$ can be quite different from that of $|L_z|=1$.  In fact, as has already been pointed out by Goryo in Ref.~\onlinecite{Goryo2008}, in the continuum limit, the skew impurity scattering diagram
for the lowest order impurity contribution to $\sigma_H(\omega)$ is nonzero only for chiral superconductors with
$|L_z|=1$ and vanishes for $|L_z|\ge 2$.  More generally, to have a non-zero $\sigma_H$ in the continuum limit, the azimuthal angular integral of $k_x k_y \Delta_1 \Delta^*_2$, where $\Delta_{1,2}$ are the two components of the chiral order parameter, must be non-zero.  While the details differ somewhat for the different mechanisms, the $k_x k_y$ in the angular integral effectively arises from the current (or velocity) operators in $\sigma_{xy}$ and $\Delta_1 \Delta^*_2$ is the lowest order contribution that directly brings in the chirality to which $\sigma_H$ is proportional.
  It follows that $\sigma_H\ne 0$ only for $|L_z|=1$.  The vanishing of $\sigma_H$ for higher chirality superconductors in the continuum limit
is a concern for $\mathrm{UPt_3}$ because the observed Kerr signal in $\mathrm{UPt_3}$~\cite{Schemm2014} is actually larger than in
$\mathrm{Sr_2 Ru O_4}$~\cite{Xia2006}. To get a nonzero anomalous Hall conductivity for $\mathrm{UPt_3}$ from chiral $f$-wave order, one needs to include lattice or bandstructure effects.

\begin{figure}[htp]
  \centering
      \includegraphics[width=0.45\textwidth]{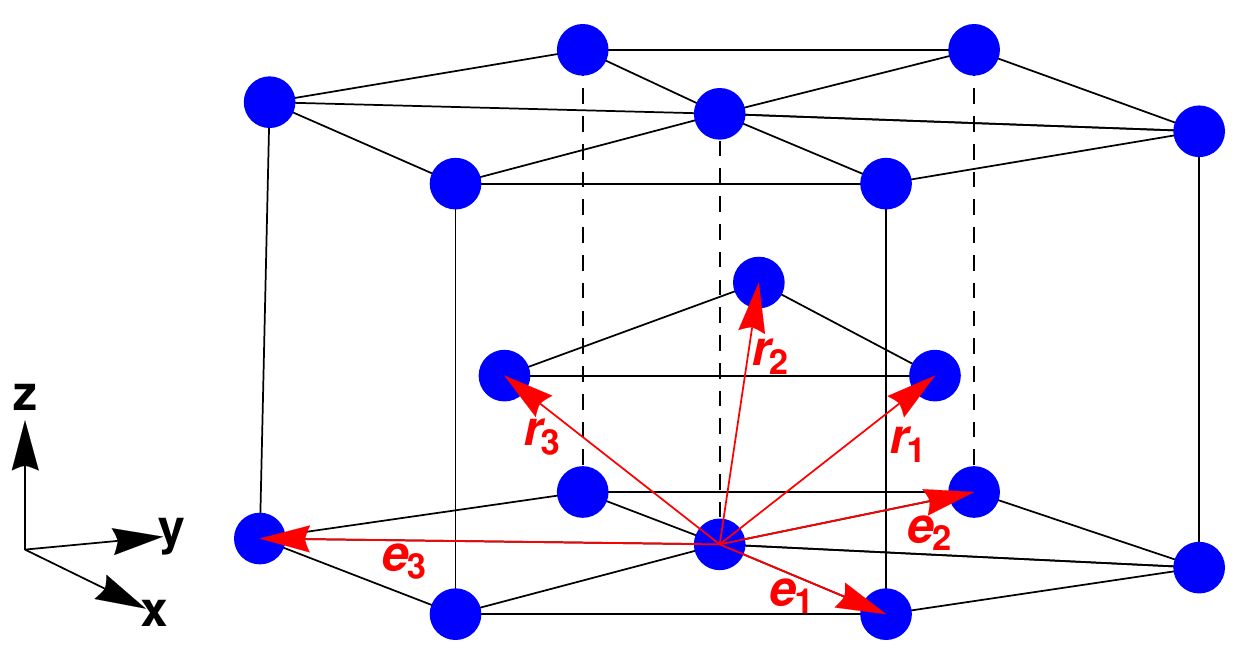}
  \caption{Crystal structure of $\mathrm{UPt_3}$. Blue disks denote the positions of $\mathrm{U}$ atoms. There is a $\mathrm{Pt}$ atom (not shown) between each nearest neighbor intra-layer pair of $\mathrm{U}$ atoms.  The vectors $\vec{e}_i$ and $\vec{r}_i$ connect
  two nearest neighbor intra-layer and inter-layer $\mathrm{U}$ atoms, respectively. The coordinate system is chosen such that $\hat{x}\parallel \vec{e}_1$. }
  \label{fig:lattice}
\end{figure}

$\mathrm{UPt_3}$ exhibits multiple superconducting phases in its temperature-magnetic field phase diagram~\cite{Fisher1989,Bruls1990,
Adenwalla1990}.  At zero field it undergoes two separate superconducting transitions at $T_c^+\approx 0.55 K$ and
$T_c^- \approx 0.5 K$~\cite{Joynt2002,Gouchi2015,Hayden1992,Lussier1996,AForder}. A nonzero Kerr rotation~\cite{Schemm2014} has been observed only in the superconducting phase below $T_c^-$.
To study whether this $\mathrm{UPt_3}$ Kerr effect can arise from the multi-band mechanism, one needs a model with at least two bands. The simplest case is two bands arising from the $\mathrm{ABAB}$ stacking of the hexagonal planes of the $\mathrm{U}$ atoms along the crystal $c-$axis. (See Fig.~\ref{fig:lattice}.)  Due to this stacking, the crystal has a close-packed
hexagonal lattice structure corresponding to the nonsymmorphic space group $P6_3/mmc$. One can ask if the two bands
resulting from this stacking can give rise to a nonzero Kerr effect. In fact, as will be discussed later, one can show that a simple chiral $d$- or $f$-wave pairing on a triangular
lattice with $\mathrm{ABAB}$ stacking gives zero, even including lattice effects beyond the continuum
limit.

Recently, Yanase~\cite{Yanase2016} argued that, due to the nonsymmorphic space group, the spin triplet superconducting order parameter is not a simple
chiral $f$-wave or a combination of only $f$- and $p$-wave.
Chiral $d$-wave pairing also mixes with the symmetry of the $E_{2u}$ representation of the crystal lattice point group $D_{6h}$.
In this model,  chiral $f$- and $p$-wave are even in the sublattice index, which can be thought of as an extra pseudospin index, while chiral $d$-wave is odd in that index and, consequently, chiral $f$-pairing is a
triplet in the $\mathrm{AB}$-sublattice subspace while the chiral $d$-wave pairing is a singlet.
Both $f$- and $d$-components involve nearest-neighbor interlayer
pairing and are of the same magnitude, while the chiral $p$-wave component involves pairing within the basal plane and is expected to be
smaller. The smaller $p$-wave pairing amplitude is presumably conjectured because of the relatively larger in-plane $\mathrm{U-U}$ atom distance~\cite{UPt3crystal} and perhaps also because the chiral $p$-component is energetically unfavorable since it pairs only one spin component. The mixing of chiral $f$- and $d$-wave leads to a more complex chiral $f+d$ pairing order parameter that is nonunitary.\cite{Yanase2016}

As a simple model, following Yanase, we study the two bands, resulting from the $\mathrm{ABAB}$ stacking, that model the ``starfish" like Fermi surfaces~\cite{Joynt2002,McMullan2008},
centered on the $A$ point at the top and bottom of the Brillouion zone (BZ). There are also four other Fermi
surface sheets resolved experimentally~\cite{Joynt2002,McMullan2008}, which, however, will not be considered in this paper.
The four other Fermi sheets are not simply related by stacking since, in general, the two bands due to the stacking (the bonding and anti-bonding bands) are well separated in energy and only one of them crosses the Fermi energy.
However, in the case of the ``starfish"  Fermi surfaces on the BZ boundary, without spin-orbit coupling (SOC) the two bands are degenerate by symmetry on the top and bottom BZ faces. With SOC, band degeneracies remain along six directions on the top and bottom surfaces.  These bands give a particularly simple two-band model for studying the intrinsic multiband mechanism of the Kerr effect.

In this paper we show that this two band model with a mixed $f+d$ wave superconducting order parameter can give rise to a nonzero Kerr effect with or without the small chiral $p$-wave pairing component. We find that mixing of the chiral $d$-component with the chiral $f$-wave pairing is essential for a nonzero $\sigma_H$. We also find that the nature of the terms that contribute to $\sigma_H$ are distinct from the terms that give a nonzero contribution for the $\mathrm{Sr_2RuO_4}$ case~\cite{Taylor2012}. From $\sigma_H$ we estimate the Kerr angle and find it to be about 10\% of the experimental value in $\mathrm{UPt_3}$~\cite{Schemm2014}.  Factors that might increase (or decrease) this estimate are discussed.

Although our work is not a complete theory of the Kerr effect for $\mathrm{UPt_3}$,  it captures a key possible contribution and more generally illustrates the necessary ingredients  for a non-zero Kerr effect for a higher chirality superconductor, a case which is noticeably more subtle than that of chiral $p$-wave.

The paper is organized as follows. In Sec.~\ref{sec:model} we describe the BdG Hamiltonian that we use for the starfish-like Fermi surface.
In Sec.~\ref{sec:sigmaH} we derive an approximate expression for $\sigma_H(\omega)$ for this BdG Hamiltonian, evaluate it numerically, and identify the key
ingredients of the result. The estimation of the Kerr angle from $\sigma_H$ and comparison to experiment
are given in Sec.~\ref{sec:Kerrangle}. Sec.~\ref{sec:conclusion} contains our conclusions and further discussions. Some technical computational details are relegated
to the Appendices.

\section{Model}~\label{sec:model}

We focus on a two-band model proposed by Yanase~\cite{Yanase2016} to describe the starfish Fermi surface (FS) of UPt$_3$.  With the two sublattices and two spin components, the BdG Hamiltonian can be written in terms of an eight component spinor $\Psi(k)$ whose transpose is defined as
\begin{gather}
\Psi^T_{\vec{k}} \equiv (c_{\vec{k}1\uparrow},  c_{\vec{k}2\uparrow} , c_{\vec{k}1\downarrow} , c_{\vec{k}2\downarrow} , c_{-\vec{k}1\uparrow}^\dagger
, c_{-\vec{k}2\uparrow}^\dagger , c_{-\vec{k}1\downarrow}^\dagger , c_{-\vec{k}2\downarrow}^\dagger),
\end{gather}
where $c_{\vec{k}is}$ is the anihilation operator for an electron with momentum $\vec{k}$, sublattice index $i$ and spin
quantum number $s$. In this basis the BdG Hamiltonian can be written as
\begin{gather}
\mathcal{H}_{\mathrm{BdG}}=\frac{1}{2}\sum_{\vec{k}\in \mathrm{BZ}} \Psi^\dagger_{\vec{k}} \hat{\mathcal{H}}_{\mathrm{BdG}}(\vec{k}) \Psi_{\vec{k}},
\end{gather}
with
\begin{gather}
\hat{\mathcal{H}}_{\mathrm{BdG}}(\vec{k}) =
\begin{pmatrix}
\hat{\mathcal{E}}(\vec{k}) & \hat{\Delta}(\vec{k}) \\
\hat{\Delta}^\dagger(\vec{k}) & -\hat{\mathcal{E}}^T(-\vec{k})
\end{pmatrix}, \label{eq:HBdG}
\end{gather}
where $\hat{\mathcal{E}}(\vec{k})$ is the normal state Hamiltonian and $\hat{\Delta}_{\vec{k}}$ is the superconducting order parameter, both $4\times 4$ matrices.

\subsection{Normal state Hamiltonian and Fermi surfaces}
Using $\sigma_{\alpha}$ and $s_{\alpha}$ to denote the four Pauli matrices for the two sublattices and spin, respectively, we can write the normal state Hamiltonian $\hat{\mathcal{E}}(\vec{k})$ as
\begin{gather}
\hat{\mathcal{E}}(\vec{k})=\xi_{\vec{k}} \sigma_0  s_0 + \frac{\epsilon_{\vec{k}}}{\sqrt{2}} \; \sigma_{+} s_0 + \frac{\epsilon_{\vec{k}}^*}{\sqrt{2}} \; \sigma_{-} s_0 + \vec{g}_{\vec{k}}\cdot \vec{s} \, \sigma_3,
\end{gather}
where $\sigma_{\pm }=(\sigma_1 \pm i \sigma_2)/\sqrt{2}$ and $\xi_{\vec{k}},\epsilon_{\vec{k}}$ and $\vec{g}_{\vec{k}}$ are given by
\begin{subequations}
\begin{align}
\xi_{\vec{k}} & =   2 t \sum_{i=1}^3 \cos \vec{k}_{\parallel}\cdot\vec{e}_i + 2 t_z \cos k_z -\mu, \\
\epsilon_{\vec{k}} & = 2 t^\prime \cos \frac{k_z}{2} \sum_{i=1}^3 e^{i\vec{k}_{\parallel}\cdot \vec{r}_i}, \\
\vec{g}_{\vec{k}} & = \hat{z} \;  \alpha \sum_{i=1}^3 \sin \vec{k}_{\parallel}\cdot \vec{e}_i.
\end{align}
\end{subequations}
Here, $\xi_{\vec{k}}$ contains all nearest neighbor (NN) hoppings within the same sublattice, both in-plane hopping with parameter $t$ and intra-sublattice NN hopping along the $c$-axis with parameter $t_z$, $\mu$ is the chemical potential and $\vec{k}_{\parallel}=(k_x,k_y,0)$.  The three unit vectors, $\vec{e}_i=(\cos\phi_i,\sin \phi_i ,0)$ with $\phi_i=(i-1)\frac{2\pi}{3}$ and $i=\{1,2,3\}$, are defined within the plane as shown in Fig.~\ref{fig:lattice}. (All lattice spacings are set to unity.)  $\epsilon_{\vec{k}}$ describes inter-sublattice NN hopping with parameter $t^\prime$.
The prefactor $\cos \frac{k_z}{2}$ in $\epsilon_{\vec{k}}$ comes from the fact that these hoppings are defined on the inter-sublattice bonds which are
described by three nonprimitive lattice vectors: $\vec{r}_i=(\frac{1}{\sqrt{3}} \cos \phi_i^\prime,
\frac{1}{\sqrt{3}}\sin \phi_i^\prime, \frac{1}{2})$, with $\phi_i^\prime=\frac{\pi}{6}+(i-1)\frac{2\pi}{3}$. $\vec{g}_k\cdot \vec{s}$ is a Kane-Mele type spin orbit coupling (SOC)
~\cite{Fischer2011,Maruyama2012} that is allowed since the local symmetry of each $\mathrm{U}$ atom is $D_{3h}$, which does not have inversion.
Note this SOC term cannot exist between two different sublattices
because the center of the inter-sublattice $\mathrm{U-U}$ bond is inversion symmetric. Also the SOCs for the two sublattices must have
opposite signs in order for the $\mathrm{U}$ lattice to restore its global $D_{6h}$ symmetry which preserves inversion~\cite{Yanase2016}. This explains the presence
of the Pauli matrix $\sigma_3$ in the SOC term in the expression of $\hat{\mathcal{E}}(\vec{k})$. The parameter $\alpha$ in $\vec{g}_{\vec{k}}$ characterizes the SOC
strength.

\begin{figure*}[tp]
\centering
\begin{tabular}{c @{\hspace{4cm}} c}
\hspace{-1.5cm}
\includegraphics[width=0.33\linewidth]{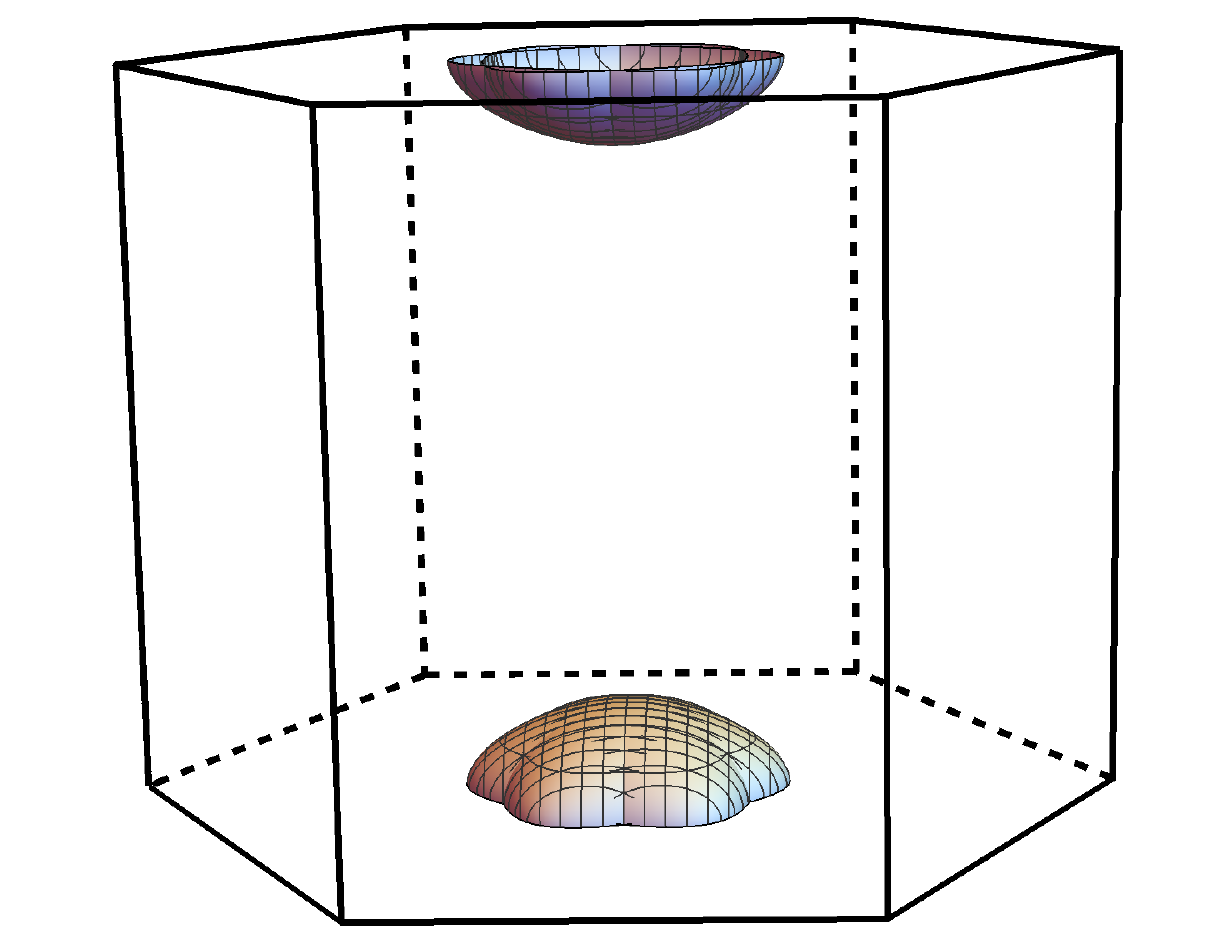} &
\hspace{-1cm}
\includegraphics[width=0.26\linewidth]{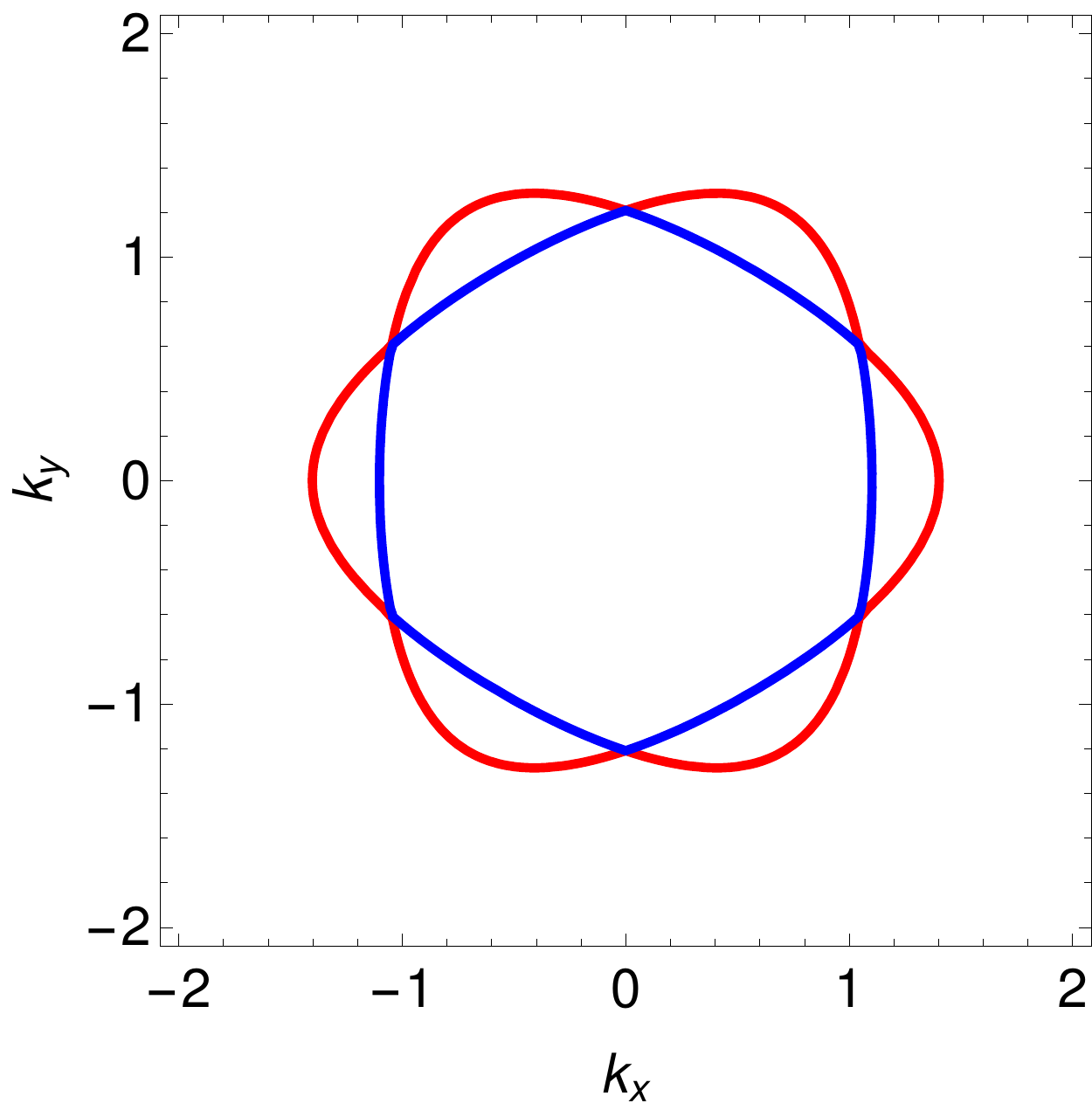} \\
\small  \hspace{-1cm} $(a)$  &  \hspace{-0.45 cm}$(b)$
\end{tabular}
\caption{Starfish Fermi surface (FS). $(a)$ FS in the three dimensional Brillouion zone of $\mathrm{UPt_3}$; $(b)$ FS contours in the plane of $k_z=\pi$. The red (blue) line is the $E_{+}^{(n)}(\vec{k})=0$ ($E_{-}^{(n)}(\vec{k})=0$) constant energy contour. Parameters used are $(t,t_z,t^\prime,\alpha,\mu)=(1,-4,1,2,12)$. }
\label{fig:YAFS}
\end{figure*}

Diagonalizing the Hamiltonian $\hat{\mathcal{E}}(\vec{k})$ gives the two normal state band dispersions,
$E^{(n)}_{\pm}(\vec{k})=\xi_{\vec{k}}\pm \sqrt{g_{\vec{k}}^2+|\epsilon_{\vec{k}}|^2}$, each of which is two-fold degenerate. The Fermi surfaces are shown in Fig.~\ref{fig:YAFS} for the parameters $(t,t_z,t^\prime,\alpha,\mu)=(1,-4,1,2,12)$ from Ref.~\onlinecite{Yanase2016}.
Fig.~\ref{fig:YAFS}(a) shows that the FS is centered around the $A-$point of the BZ; while Fig.~\ref{fig:YAFS}(b) presents a cut of the FS on the zone boundary
$k_z=\pi$ plane. Note, from Fig.~\ref{fig:YAFS}(b), the two Fermi surfaces intersect at six points on that plane since $\epsilon_{\vec{k}}=0$ for $k_z=\pi$ and $g_{\vec{k}}$ vanishes along the six-fold symmetric directions: $k_y/k_x=\tan \theta_i $ with
$\theta_i=\frac{\pi}{6}+(i-1)\frac{\pi}{3}$.

\subsection{Superconducting order parameter $\hat{\Delta}(\vec{k})$}

The superconducting order parameter $\hat{\Delta}(\vec{k})$ proposed in Ref.~\onlinecite{Yanase2016} is an  $E_{2u}$ state that can be written as
$\hat{\Delta}(\vec{k})=\eta_1 \hat{\Gamma}_1(\vec{k}) + \eta_2 \hat{\Gamma}_2(\vec{k})$. Here $\hat{\Gamma}_1(\vec{k})$ and $\hat{\Gamma}_2(\vec{k})$ are two basis
functions of the $E_{2u}$ representation, and $(\eta_1,\eta_2)=\Delta_0 (1,i \eta)/\sqrt{1+\eta^2}$, with overall pairing magnitude $\Delta_0$  and $\eta$ a real number that controls the anisotropy of the order parameter.
Due to the relative phase between $\eta_1$ and $\eta_2$, $\hat{\Delta}(\vec{k})$ is chiral, with the chirality  determined by the sign of $\eta$.

$\hat{\Gamma}_1(\vec{k})$ and $\hat{\Gamma}_2(\vec{k})$ are both triplets in spin as suggested by experiments~\cite{Sauls1994,Tou1998,Joynt2002}. The spatial parts of $\hat{\Gamma}_1(\vec{k})$ and $\hat{\Gamma}_2(\vec{k})$  contain not only
$f$- and $p$-wave components but also a $d$-wave component as discussed above.
Spatial inversion operation not only transforms $\vec{k}\rightarrow -\vec{k}$ but also interchanges the two sublattices.
The $f$- and $p$-wave components are odd functions of $\vec{k}$ and triplets in the sublattice index, while the $d$-component
is an even function of $\vec{k}$ but a sublattice singlet.
As mentioned above, the pairing amplitudes of the $f$- and $d$-wave components connect different sublattices while the $p$-wave component pairs sites on the same sublattice.
The $f$- and $d$-components are of similar magnitude while the $p$-wave is smaller. In the following, we will ignore this small $p$-component.
Then the two basis functions $\hat{\Gamma}_1$ and $\hat{\Gamma}_2$ can be written as~\cite{Yanase2016}
\begin{subequations}
\begin{align} \label{eq:OPG1G2}
\hat{\Gamma}_1(\vec{k}) & = \big\{ f_{(x^2-y^2)z}(\vec{k})\sigma_1 - d_{yz}(\vec{k}) \sigma_2 \big\} s_1, \\
\hat{\Gamma}_2(\vec{k}) & = \big\{ f_{xyz}(\vec{k})\sigma_1 - d_{xz}(\vec{k}) \sigma_2 \big\} s_1,
\end{align}
\end{subequations}
where, for nearest-neighbor intersublattice pairing,
\begin{subequations}
\begin{align}
f_{(x^2-y^2)z}(\vec{k}) & = -\sin \frac{k_z}{2}[\cos \frac{k_x}{2}\cos\frac{k_y}{2\sqrt{3}}-\cos \frac{k_y}{\sqrt{3}}], \\
f_{xyz}(\vec{k}) & =\sqrt{3} \sin \frac{k_x}{2} \sin \frac{k_y}{2\sqrt{3}} \sin \frac{k_z}{2}, \\
d_{yz}(\vec{k}) & = -\sin \frac{k_z}{2}[\cos \frac{k_x}{2} \sin \frac{k_y}{2\sqrt{3}}+\sin \frac{k_y}{\sqrt{3}}], \\
d_{xz}(\vec{k}) & = -\sqrt{3} \sin \frac{k_x}{2} \cos \frac{k_y}{2\sqrt{3}} \sin\frac{k_z}{2}.
\end{align}
\end{subequations}
In the expressions for $\hat{\Gamma}_1(\vec{k})$ and $\hat{\Gamma}_2(\vec{k})$, the spin Pauli matrix $s_1=s_3\, i s_2$ indicates that the spin triplet pairing $\vec{d}$ vector is along the $\hat{z}$-direction (or the crystal $c$-axis).
The presence of sublattice Pauli matrices $\sigma_1$ and $\sigma_2$ comes from the fact that the $f$- and $d$-wave components are derived from the real
and imaginary part, respectively, of a pairing amplitude for electrons from NN inter-sublattice $\mathrm{U}$ ions. Because of the mixing
between the $f$- and $d$-wave components,
\begin{align}
\hat{\Delta}(\vec{k})\hat{\Delta}^\dagger(\vec{k}) & = \big\{ |f_{\vec{k}}|^2  +|d_{\vec{k}}|^2 \big\} \sigma_0 s_0  - i \big\{ f_{\vec{k}} d^*_{\vec{k}}-f_{\vec{k}}^* d_{\vec{k}} \big\}
\sigma_3 s_0 \label{eq:DeltaDelta}
\end{align}
has a term which is not proportional to the identity matrix $\sigma_0 s_0$, which makes $\hat{\Delta}(\vec{k})$ nonunitary~\cite{Sigrist1991}. In Eq.~\eqref{eq:DeltaDelta},
\begin{subequations}
\begin{align}
f_{\vec{k}} & \equiv \eta_1 f_{(x^2-y^2)z} (\vec{k})+\eta_2  f_{xyz}(\vec{k}), \\
d_{\vec{k}} & \equiv \eta_1 d_{yz} (\vec{k})+\eta_2 \, d_{xz}(\vec{k}).
\end{align}
\end{subequations}

\subsection{Reduction of the BdG Hamiltonian}

The expressions for $\hat{\mathcal{E}}(\vec{k})$ and $\hat{\Delta}(\vec{k})$ defined above can now be substitued into the BdG Hamiltonian given by Eq.~\eqref{eq:HBdG}. One finds $\mathcal{H}_{\mathrm{BdG}}(\vec{k})$ reduces to two decoupled $4\times 4$ blocks:
\begin{align} \label{eq:HBdG2}
\mathcal{H}_{\mathrm{BdG}} & =\mathcal{H}^{(a)}+\mathcal{H}^{(b)} \nonumber \\
& = \frac{1}{2} \sum_{i=a,b} \sum_{\vec{k}\in \mathrm{BZ}} [\Psi^{(i)}_{\vec{k}}]^\dagger \hat{\mathcal{H}}^{(i)}(\vec{k})
\Psi_{\vec{k}}^{(i)},
\end{align}
with
\begin{subequations}
\begin{gather}
\hspace{-0.25cm}
\hat{\mathcal{H}}^{(a)}=
\begin{pmatrix}
  \xi_{\vec{k}} +g_{\vec{k}}   &  \epsilon_{\vec{k}}         & 0                           & \Delta_{12}(\vec{k}) \\
  \epsilon_{\vec{k}}^*         & \xi_{\vec{k}}-g_{\vec{k}}   &  \Delta_{21}(\vec{k})       & 0\\
   0                           & \Delta_{21}^*(\vec{k})      & -\xi_{\vec{k}}-g_{\vec{k}}  &  -\epsilon_{\vec{k}} \\
  \Delta_{12}^*(\vec{k})       & 0                           & -\epsilon_{\vec{k}}^*       &  -\xi_{\vec{k}}+g_{\vec{k}}
\end{pmatrix},  \label{eq:Hblock1} \\
\hspace{-0.25cm}
\hat{\mathcal{H}}^{(b)}=
\begin{pmatrix}
  \xi_{\vec{k}} - g_{\vec{k}}   &  \epsilon_{\vec{k}}         & 0                           & \Delta_{12}(\vec{k}) \\
  \epsilon_{\vec{k}}^*         & \xi_{\vec{k}} + g_{\vec{k}}   &  \Delta_{21}(\vec{k})       & 0\\
   0                           & \Delta_{21}^*(\vec{k})      & -\xi_{\vec{k}} + g_{\vec{k}}  &  -\epsilon_{\vec{k}} \\
  \Delta_{12}^*(\vec{k})       & 0                           & -\epsilon_{\vec{k}}^*       &  -\xi_{\vec{k}} - g_{\vec{k}}
\end{pmatrix}. \label{eq:Hblock2}
\end{gather}
\end{subequations}
The two bases are
\begin{subequations}
\begin{gather}
\Psi_{\vec{k}}^{(a)}=
\begin{pmatrix}
c_{\vec{k} 1 \uparrow} &
c_{\vec{k} 2 \uparrow} &
c^\dagger_{-\vec{k} 1 \downarrow} &
c^\dagger_{-\vec{k} 2 \downarrow}
 \end{pmatrix}^{T},
\\
\Psi_{\vec{k}}^{(b)}=
\begin{pmatrix}
c_{\vec{k} 1 \downarrow} &
c_{\vec{k} 2 \downarrow} &
c^\dagger_{-\vec{k} 1 \uparrow} &
c^\dagger_{-\vec{k} 2 \uparrow}
\end{pmatrix}^{T}.\label{eq:PsiaPsib}
\end{gather}
\end{subequations}
In the above equations, $g_{\vec{k}}\equiv \hat{z}\cdot \vec{g}(\vec{k})$, $\Delta_{12}(\vec{k})  \equiv f_{\vec{k}} + i \, d_{\vec{k}}$ and
$ \Delta_{21}(\vec{k}) \equiv f_{\vec{k}} - i \, d_{\vec{k}}$, where $1,2$ are sublattice labels.
The two blocks are connected to each other
by spin inversion, $\uparrow \leftrightarrow \downarrow$, which leaves all matrix elements of $\hat{\mathcal{H}}^{(a)}(\vec{k})$
and $\hat{\mathcal{H}}^{(b)}(\vec{k})$ unchanged except for a change in the sign of the SOC term, $g_{\vec{k}}$. However, as will be shown later, the Hall conductivity $\sigma_H(\omega)$
is an even function of $g_{\vec{k}}$. Therefore we only need to focus on one block, say $\hat{\mathcal{H}}^{(a)}(\vec{k})$, and multiply the $\sigma_H$ computed for that block by a factor of two.  An additional factor of 1/2, arising from the double-counting of degrees of freedom in BdG theory, will cancel this factor of 2. Hereafter, we drop the superscript $(a)$ in $\hat{\mathcal{H}}^{(a)}(\vec{k})$ and simply denote it as $\hat{\mathcal{H}}(\vec{k})$ for brevity.  Note that this decomposition into two $4\times 4$ blocks is only possible in the absence of the intralayer $p$-wave pairing.

\begin{figure}[tp]
\centering
\includegraphics[scale=0.44]{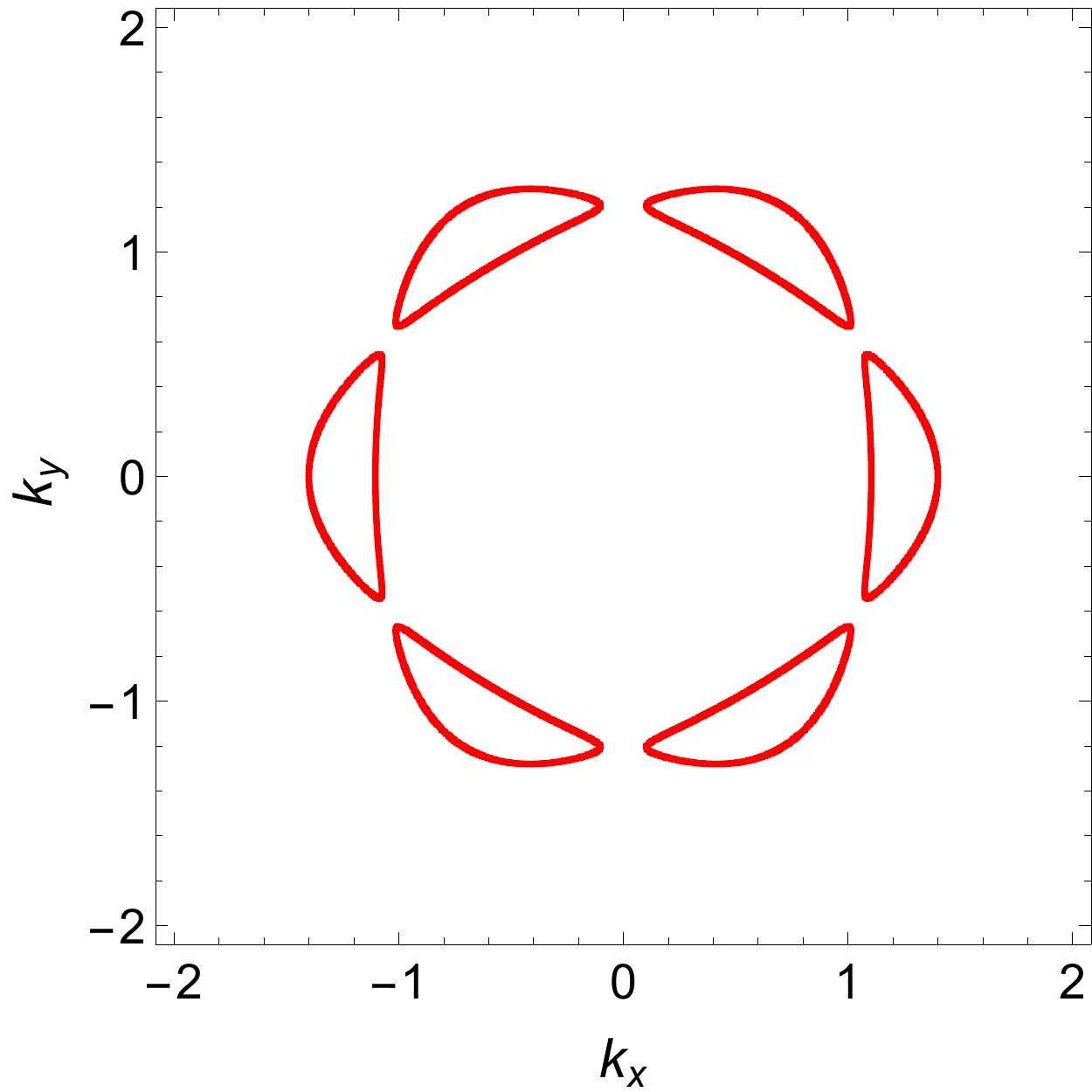}
\caption{Bogoliubov quasiparticle energy line nodes of the BdG Hamiltonian $\hat{\mathcal{H}}(\vec{k})$ at $k_z=\pi$.
The parameter $\Delta_0=0.1\,t$. Other parameters used are the same as in Fig.~\ref{fig:YAFS}.}
\label{fig:YAscnode}
\end{figure}

From $\hat{\mathcal{H}}(\vec{k})$ one can obtain the Bogoliubov quasiparticle energies, which have
line nodes on the $k_z=\pm \pi$ plane that form six rings, as shown in Fig.~\ref{fig:YAscnode}. These nodal rings are counter examples to Blount's
theorem~\cite{Blount1985,Norman1995,Micklitz2009,Kobayashi2016,Micklitz2017,Micklitz2017b} and are topologically protected as a joint consequence of both the non-symmorphic group symmetries and the nonzero spin orbital coupling, as discussed in
Refs.~\onlinecite{Kobayashi2016,Micklitz2017,Micklitz2017b}.

\section{Computation of the anomalous Hall conductivity $\sigma_H(\omega)$}~\label{sec:sigmaH}

The Hall conductivity $\sigma_H(\omega)$ can be computed from the Kubo formula~\cite{Mahan1990,Taylor2012}
\begin{gather}
\sigma_H(\omega) = \frac{i}{2\omega} \lim_{\vec{q}\rightarrow 0} \big\{ \pi_{xy}(\vec{q},\omega)-\pi_{yx}(\vec{q},\omega)\big\}, \label{eq:sigmadef}
\end{gather}
where $\pi_{xy}(\vec{q},\omega)$ is the electric current density $\hat{J}_x$-$\hat{J}_y$ correlator. At the one-loop level $\pi_{xy}$
is given by (setting $e=\hbar=c=1$)
\begin{widetext}
\begin{gather}
 \pi_{xy}(\vec{q}=0,i\nu_m) = \sum_{\vec{k}} T \sum_{n} \mathrm{Tr}\big\{\hat{v}^x(\vec{k}) \hat{G}(\vec{k},i \omega_n+i\nu_m)\hat{v}^y(\vec{k}) \hat{G}(\vec{k},i\omega_n)\big\}, \label{eq:pixy}
\end{gather}
\end{widetext}
where $T$ is the temperature (set to $T=0$ at the end of the calculation) and  $\omega_n=(2n+1)\pi T$ and $\nu_m= 2 m \pi T$ are fermionic and bosonic Matsubara frequencies, respectively.
$\hat{G}(\vec{k},i\omega_n)$ is the Green's function of the $4\times 4$ block Hamiltonian $\hat{\mathcal{H}}(\vec{k})$ with inverse defined by
\begin{gather}
\hat{G}^{-1}(\vec{k},i\omega_n)= i\omega_n - \hat{\mathcal{H}}(\vec{k}). \label{eq:Ginv1}
\end{gather}
From $\mathrm{det}\, \hat{G}^{-1}(\vec{k},i\omega_n)=0$ one obtains the Bogoliubov quasiparticles energies of the Hamiltonian $\hat{\mathcal{H}}(\vec{k})$.
However, the equation to be solved is not a quadratic equation for $\omega_n^2$ but a quartic equation in $\omega_n$ (see Eq.~\eqref{eq:quartic} in
App.~\ref{sec:appendixFullG}).  Consequently, the analytic expressions for the quasiparticle energies as well as the final expression for $\sigma_H(\omega)$ are quite lengthy,
and these results are summarized in App.~\ref{sec:appendixFullG} in Eqs.~\eqref{eq:appendixSigmaH} to \eqref{eq:Oxy2def}. From these expressions it is difficult to identify
which ingredients are essential to obtain
a nonzero $\sigma_H(\omega)$, and so we also compute $\sigma_H$ perturbatively to obtain a much simpler expression that is valid at intermediate to high frequencies.

We  treat the $d$-wave component  of the superconducting order parameter
as a perturbation and write $\hat{\mathcal{H}}(\vec{k})=\hat{\mathcal{H}}_0(\vec{k})+\hat{\mathcal{H}}^\prime(\vec{k})$
where
\begin{gather}
\hat{\mathcal{H}}_0
=\begin{pmatrix}
\xi_{\vec{k}} + g_{\vec{k}} & \epsilon_{\vec{k}}            & 0                            & f_{\vec{k}} \\
\epsilon_{\vec{k}}^*        & \xi_{\vec{k}} - g_{\vec{k}}   & f_{\vec{k}}                   & 0 \\
0                           & f^*_{\vec{k}}                  & -\xi_{\vec{k}} - g_{\vec{k}} & -\epsilon_{\vec{k}} \\
f^*_{\vec{k}}                & 0                             & -\epsilon^*_{\vec{k}}                & -\xi_{\vec{k}} + g_{\vec{k}}
 \end{pmatrix},
\end{gather}
and
\begin{gather}
 \hat{\mathcal{H}}^\prime =
\begin{pmatrix}
0 & 0 & 0 & i d_{\vec{k}} \\
0 & 0 &  - i d_{\vec{k}} & 0 \\
0 & i d^*_{\vec{k}} &  0 \\
-i d^*_{\vec{k}} & 0 & 0 & 0
\end{pmatrix}.
\end{gather}
$\hat{\mathcal{H}}_0$ and $\hat{\mathcal{H}}^\prime$ will be taken as the ``unperturbed" and ``perturbed" Hamiltonian, respectively.
We choose this particular partition because it is precisely the $d-$component superconducting order parameter part that makes the Bogoliubov quasiparticle
energy expression complicated (see Eq.~\eqref{eq:quartic} in App.~\ref{sec:appendixFullG}) and also because, as we will see later, the leading order contribution to $\sigma_H$ is linear in $d_{\vec{k}}$.

Since we are including the effect of $\hat{\mathcal{H}}^\prime$ only perturbatively, the results are only reliable for sufficently large $\omega$. Actually, the perturbative expansion is in $\beta_{\vec{k}} \propto i (f_{\vec{k}} d^*_{\vec{k}}-f^*_{\vec{k}}d_{\vec{k}}) g_{\vec{k}}$,
not just $d_{\vec{k}}$ (see  Eq.~\eqref{eq:quartic} in App.~\ref{sec:appendixFullG} for details).
So,  the perturbative results are reliable for $\omega \gg \beta_{\vec{k}} \sim (\Delta_0^2 \; \alpha)^{1/3}$, where $\alpha$ is the SOC strength.    The full Green's function results and the perturbative results for $\sigma_H(\omega)$, are compared in App.~\ref{sec:appendixFullG} in
Figs.~\ref{fig:compareIm} and \ref{fig:compareRe}, showing the two are essential identical beyond $\omega \gtrsim 4 t$.  Since the laser frequency at which the Kerr effect has been measured is $\omega \approx 0.8~\mathrm{eV}$~\cite{Schemm2014}, which is $> 20t$ in our model, the perturbative results can be used to compare to experiment.

\subsection{Perturbative calculation}
Here we discuss the perturbative calculation of $\sigma_H$, with further details given in App.~\ref{sec:appendixPixy}.  Quantities of different order in
$\hat{\mathcal{H}}^\prime$ are represented by superscripts $(0),(1),\cdots$.  First, consider zeroth order described by the Hamiltonian $\hat{\mathcal{H}}_0(\vec{k})$.  The Bogoliubov quasiparticle energies, $E_{\pm}$, are
\begin{align}
E_{\pm} & =  \sqrt{a \pm  \sqrt{a^2-b}} , \label{eq:Epm}
\end{align}
with
\begin{subequations}
\begin{align}
a & = \xi_{\vec{k}}^2 +g_{\vec{k}}^2 + |\epsilon_{\vec{k}}|^2+|f_{{\vec{k}}}|^2, \\
b & = (\xi_{\vec{k}}^2-g_{\vec{k}}^2 +|f_{{\vec{k}}}|^2-|\epsilon_{\vec{k}}|^2)^2+|f_{{\vec{k}}}|^2 (\epsilon_{\vec{k}}+\epsilon^*_{\vec{k}})^2,
\end{align}
\end{subequations}
which are slightly different from those of the full Hamiltonian $\hat{\mathcal{H}}(\vec{k})$. However $E_{-}$ still has nodal rings on the $k_z=\pm \pi$ plane that are
almost identical to those obtained from the full Hamiltonian, $\hat{\mathcal{H}}(\vec{k})$, plotted in Fig.~\ref{fig:YAscnode}.  These nodal rings are protected
by the non-symmorphic space group symmetry and spin-orbit coupling~\cite{Kobayashi2016,Yanase2016}.

The velocity operators, which appear in Eq.~\eqref{eq:pixy}, are defined by the normal state Hamiltonian,
$\hat{\mathcal{H}}_N(\vec{k})$, which can be written in terms of the sublattice Pauli matrices, $\sigma_\alpha$:
\begin{gather}
\hat{\mathcal{H}}_N(\vec{k})= \xi_{\vec{k}} \sigma_0 + \vec{h}\cdot \boldsymbol{\sigma}, \label{eq:Hnorm}
\end{gather}
with $\vec{h}=(\frac{\epsilon_{\vec{k}}}{\sqrt{2}} ,\frac{\epsilon^*_{\vec{k}}}{\sqrt{2}},g_{\vec{k}})$ and
$\boldsymbol{\sigma}=(\sigma_+,\sigma_-,\sigma_3)$.
Then  $\hat{v}^x=\partial_{k_x} \hat{\mathcal{H}}_N(\vec{k}) \, \tau_0$~\cite{Lutchyn2009,Taylor2012}, where $\tau_0$
is the identity matrix for the Nambu space, or written out explicitly,
\begin{gather}
\hat{v}^x=\begin{pmatrix}
\partial_{k_x} E_a(\vec{k}) & \partial_{k_x} \epsilon_{\vec{k}} & 0 & 0 \\
\partial_{k_x} \epsilon^*_{\vec{k}} & \partial_{k_x} E_b(\vec{k}) & 0 & 0 \\
0 & 0 &  \partial_{k_x} E_a(\vec{k}) & \partial_{k_x} \epsilon_{\vec{k}} \\
0 & 0 & \partial_{k_x} \epsilon^*_{\vec{k}} & \partial_{k_x} E_b(\vec{k})
\end{pmatrix}, \label{eq:vxmatrix}
\end{gather}
with $E_a(\vec{k})\equiv \xi_{\vec{k}}+g_{\vec{k}}$ and $E_b(\vec{k})\equiv \xi_{\vec{k}}-g_{\vec{k}}$.
$\hat{v}^y$ can be obtained from $\hat{v}^x$ by the substitution: $\partial_{k_x} \rightarrow \partial_{k_y}$. With $\hat{v}^x,\hat{v}^y$ and  $\hat{G}^{(0)} \equiv \big\{ i\omega_n-\hat{\mathcal{H}}_0(\vec{k}) \big\}^{-1}$, one can compute the zeroth order current-current correlator $\pi_{xy}^{(0)}(i\nu_m)$ from Eq.~\eqref{eq:pixy}.
 However, a direct computation shows that $\pi_{xy}^{(0)}(i\nu_m)-\pi_{yx}^{(0)}(i\nu_m)=0$, so that $\sigma_H^{(0)}(\omega)\equiv 0$. In other words, a chiral $f$-wave superconducting order parameter alone does not give rise to a non-zero anomalous Hall conductivity from the multiband mechanism if the two bands arise from $\mathrm{ABAB}$ stacking.  The mixing between $f$-wave and $d$-wave components is crucial for a nonzero $\sigma_H$ and one needs to go to first order to calculate a non-zero $\sigma_H(\omega)$.

From the full Green's function $\hat{G}=\hat{G}^{(0)}+\hat{G}^{(0)} \hat{\mathcal{H}}^\prime \hat{G}^{(0)}+\cdots$,
one can define the first order Green's function as $\hat{G}^{(1)}=\hat{G}^{(0)} \hat{\mathcal{H}}^\prime \hat{G}^{(0)}$ and, from Eq.~\eqref{eq:pixy}, the first order current-current correlator is
\begin{widetext}
\begin{align}
\pi_{xy}^{(1)}(i\nu_m) & = \sum_{\vec{k}} T \sum_{n} \bigg\{ \mathrm{Tr}[\hat{v}^x \hat{G}^{(0)}(\vec{k}, i \omega_n+i\nu_m) \hat{v}^y \hat{G}^{(1)}(\vec{k}, i\omega_n)]
+  \big\{ (0) \leftrightarrow (1) \big\} \bigg\}.  \label{eq:pixy1}
\end{align}
\end{widetext}
This (or, more precisely, $\pi_{xy}^{(1)}(i\nu_m)-\pi_{yx}^{(1)}(i\nu_m)$) is evaluated in App.~\ref{sec:appendixPixy} by first writing the velocity operators and Green's functions as linear combinations of Pauli matrices to simplify computing the trace and then doing the Matsubara sum.  After
performing a Wick rotation, $i\nu_m \rightarrow \omega+i\delta$, one obtains the final expression for the Hall conductivity,
\begin{widetext}
\begin{align}
\sigma_H^{(1)}(\omega) & = \sum_{\vec{k}} 4 i [ f_{\vec{k}} d^*_{\vec{k}} - f^*_{\vec{k}} d_{\vec{k}}] \; \xi_{\vec{k}}
\bigg\{ 8  i\; g_k \; \vec{h}\cdot\partial_{k_x} \vec{h}\times \partial_{k_y} \vec{h}  \; \frac{ S_{\vec{k}}(\omega)}{\omega}
+ \Omega_{xy} \, \frac{ T_\vec{k}(\omega)}{\omega} \bigg\},\label{eq:sigmaH1-2}
\end{align}
\end{widetext}
where for brevity we have suppressed the infinitesimal imaginary part, $i\delta$, in $\omega+i\delta$. $\Omega_{xy}$ is an anti-symmetrized velocity factor given by
\begin{gather}
\Omega_{xy} \equiv  - i[\partial_{k_x} \epsilon_{\vec{k}} \partial_{k_y}\epsilon^*_{\vec{k}} -
\partial_{k_x} \epsilon^*_{\vec{k}} \partial_{k_y}\epsilon_{\vec{k}}].  \label{eq:Oxy}
\end{gather}
We have also introduced two frequency dependent functions in
Eq.~\eqref{eq:sigmaH1-2}, which are defined as (for details see App.~\ref{sec:appendixPixy})
\begin{align}
\frac{S_{\vec{k}}(\omega)}{\omega} & \equiv  F_1(\vec{k},\omega) -\frac{\xi_{\vec{k}}^2 - g_{\vec{k}}^2 -|\epsilon_{\vec{k}}|^2}{E_+ E_-} \; F_2(\vec{k},\omega), \label{eq:Skw}
\end{align}
and
\begin{align}
\frac{T_{\vec{k}}(\omega)}{\omega} & \equiv F_3(\vec{k}, \omega), \label{eq:Tkw}
\end{align}
with $F_1(\vec{k},\omega), F_2(\vec{k},\omega)$ and $F_3(\vec{k},\omega)$ given by
\begin{widetext}
\begin{subequations}
\begin{align}
F_1(\vec{k},\omega) & \approx  \frac{C_{++}}{\omega^2-4 E_+^2}  +  \frac{C_{--}}{\omega^2-4 E_-^2} +  \frac{C_{+-}}{\omega^2-(E_+ + E_-)^{2}}, \label{eq:F1omega} \\
F_2(\vec{k},\omega) & \approx \frac{D_{++}}{\omega^2-4 E_+^2}  +  \frac{D_{--}}{\omega^2 - 4 E_-^2}  +   \frac{D_{+-}}{\omega^2-(E_+ + E_-)^2},\label{eq:F2omega} \\
F_3(\vec{k},\omega) & \approx \frac{B_{+-}}{\omega^2-(E_+ + E_-)^2}. \label{eq:F3omega}
\end{align}
\end{subequations}
\end{widetext}
The $\approx$ sign means only the leading order terms in $f_{\vec{k}}$ and $d_{\vec{k}}$ have been kept.
There are seven frequency independent coefficients in the numerators of $F_1,F_2$ and $F_3$. Their expressions are
\begin{subequations}
\begin{align}
C_{++} & =-D_{--}=\frac{E_+}{(E_-^2 - E_+^2)^3}, \label{eq:CDcoeff1} \\
D_{++} & =-C_{--}=\frac{E_-}{(E_-^2 - E_+^2)^3},  \label{eq:CDcoeff2}\\
C_{+-} & = \frac{E_+^2 +  E_-^2}{2 E_+ E_- (E_+ + E_-)^3 (E_+ - E_-)^2}, \label{eq:CDcoeff3}\\
D_{+-} & = \frac{-1}{(E_+ + E_-)^3 (E_+ - E_-)^2}, \label{eq:CDcoeff4}\\
B_{+-} & = \frac{1}{2 E_+ E_- (E_+ + E_-)}.\label{eq:CDcoeff5}
\end{align}
\end{subequations}
The subscripts, $\{++,--,+-\}$, in these coefficients directly reflect the corresponding physical processes that they are associated with, which can be inferred from the
denominator of each term in the expressions of $F_1(\vec{k}, \omega)$, $F_2(\vec{k}, \omega)$ and $F_3(\vec{k}, \omega)$. For example, the first term in $F_1(\vec{k}, \omega)$ with
coefficient $C_{++}$ corresponds to a process where a Cooper pair, with momentum $(\vec{k}, -\vec{k})$, is
broken and a Bogoliubov quasiparticle pair with energies, $E_+(\vec{k})$ and $E_+(-\vec{k})$, are excited by the incident photon with a frequency $\omega$.
The two Bogoliubov quasiparticles have the same momentum $(\vec{k},-\vec{k})$ as the broken Cooper pair because the incident photon momentum $\vec{q} \approx 0$
relative to $\vec{k}$. Energy conservation of this process requires $\omega=E_+(\vec{k})+E_+(-\vec{k})=2 E_+$, which explains the denominator $\omega^2-(2 E_+)^2$ in
the first term in $F_1(\vec{k}, \omega)$. Other terms in $F_1(\vec{k}, \omega)$, $F_2(\vec{k}, \omega)$ and $F_3(\vec{k}, \omega)$ can be interpretated in a similar way.
Notice that in the expressions for $F_1(\vec{k}, \omega)$, $F_2(\vec{k}, \omega)$ and $F_3(\vec{k}, \omega)$ there is no term with a denominator $\omega^2-(E_+-E_-)^2$,
which would correspond to a $T>0$ process where a preexisting Bogoliubov quasiparticle with an energy $E_-$ gets excited to a higher energy level of $E_+$ by the incident
photon.

Finally, as noted below Eq.~\eqref{eq:PsiaPsib}, we can see from Eqs.~\eqref{eq:Skw}-\eqref{eq:Tkw}, that
$\sigma_H^{(1)}(\omega)$ is an even function of $g_{\vec{k}}$, since the two functions $S_{\vec{k}}(\omega)$ and $T_{\vec{k}}(\omega)$ depend on $\vec{k}$ only through
$E_{\pm}$, which are even in $g_{\vec{k}}$ (see Eq.~\eqref{eq:Epm}); $\Omega_{xy}$ does not depend on $g_{\vec{k}}$ (see Eq.~\eqref{eq:Oxy}), and the factor $g_k \; \vec{h}\cdot\partial_{k_x}
\vec{h}\times \partial_{k_y} \vec{h}$ is also even in $g_{\vec{k}}$ because the mixed product contributes one and only one $g_{\vec{k}}$ since $\vec{h}=(\epsilon_{\vec{k}}/\sqrt{2}
,\epsilon_{\vec{k}}^*/\sqrt{2},g_{\vec{k}})$.

\begin{figure*}[tp]
\centering
\begin{minipage}{0.5\textwidth}
\centering
\includegraphics[width=0.7\linewidth]{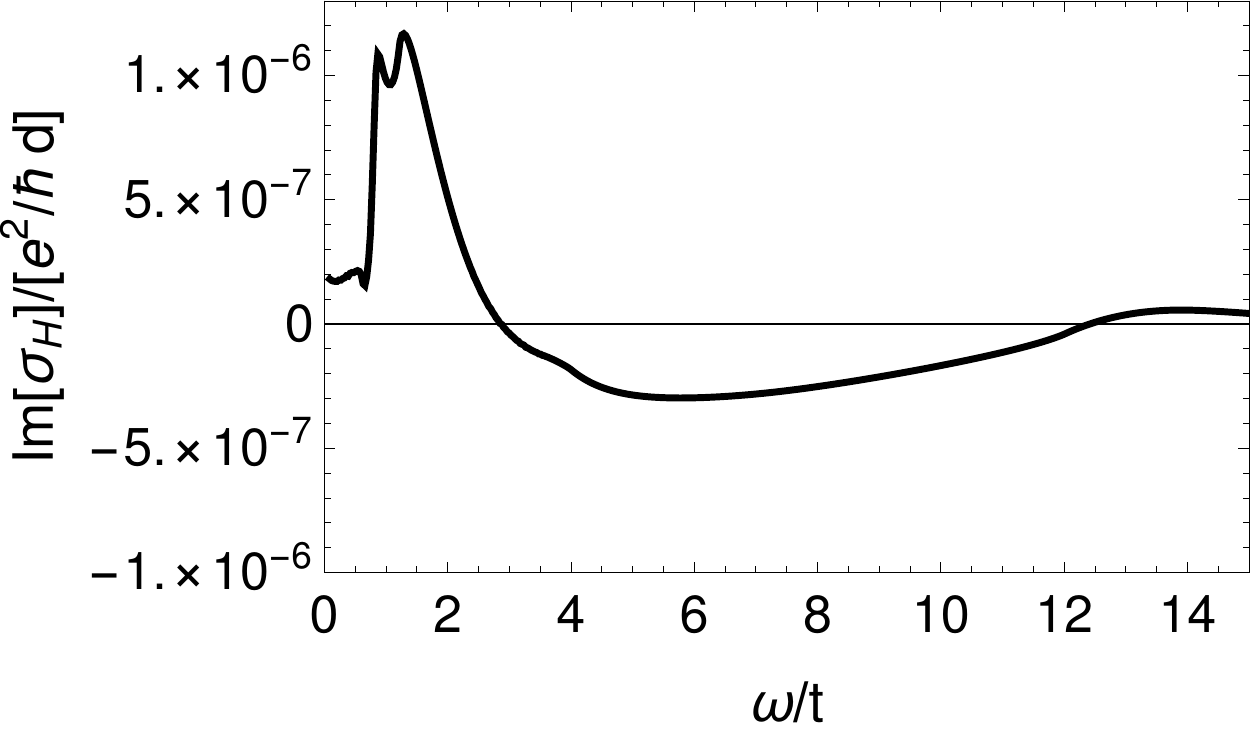}
\end{minipage}%
~
\begin{minipage}{0.5\textwidth}
\centering
\includegraphics[width=0.7\linewidth]{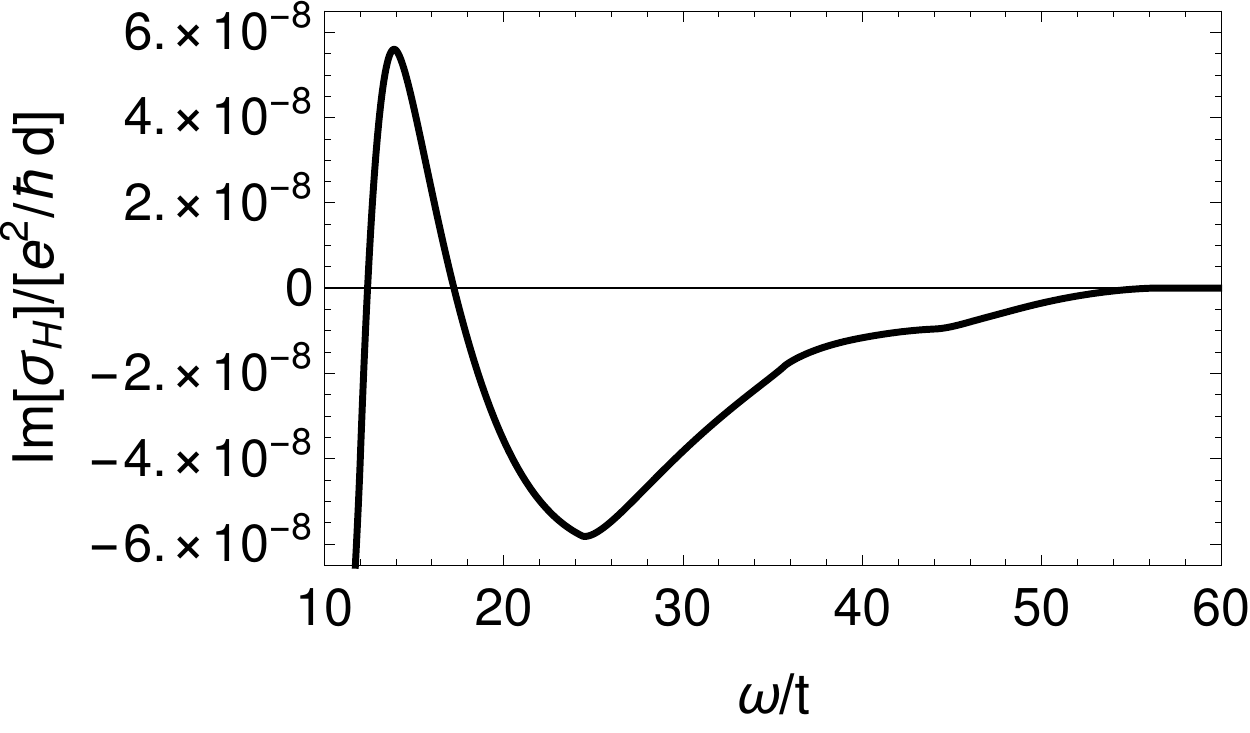}
\label{fig:ExpImL}
\end{minipage}
\caption{\label{fig:ExpIm}Numerical results for $\mathrm{Im} \; \sigma_H^{(1)}(\omega)$. Left panel: small frequency regime $\omega/t \le 14$; right
panel: large frequency regime $\omega/t \ge 10$. Note that
the vertical axis scales of the two figures are different. The unit of $\sigma_H$ is $e^2/\hbar d$, with $d$ the $\hat{c}-$axis lattice spacing of $\mathrm{UPt_3}$.
Parameters used are $(t,t_z,t^\prime,\alpha,\mu,\Delta_0,\eta)=(1,-4,1,2,12,0.1,1.0)$.}
\end{figure*}

Next we evaluate the expression for $\sigma_H^{(1)}(\omega)$ in Eq.~\eqref{eq:sigmaH1-2} numerically. Replacing $\omega$ with $\omega+i\delta$ in Eq.~\eqref{eq:sigmaH1-2}, the imaginary part can be written as
\begin{widetext}
\begin{align}
\mathrm{Im}\, \sigma_H^{(1)} & = -\frac{\pi}{2\omega}  \sum_{\vec{k}} 4 i\, [f_{\vec{k}} d^*_{\vec{k}} -f^*_{\vec{k}} d_{\vec{k}}] \; \xi_{\vec{k}}
\bigg\{ 8 i \; g_{\vec{k}} \vec{h}\cdot\partial_{k_x} \vec{h}\times \partial_{k_y} \vec{h} \,
\mathcal{A}_{1}(\vec{k},\omega)+ \Omega_{xy}\,\mathcal{A}_2(\vec{k},\omega)\bigg\} \label{eq:sigmaH1},
\end{align}
where $\mathcal{A}_1(\vec{k},\omega)$ and $\mathcal{A}_2(\vec{k},\omega)$ are:
\end{widetext}
\begin{widetext}
\begin{subequations}
\begin{align}\label{eq:A1A2}
\mathcal{A}_1(\vec{k},\omega) & \equiv [C_{++}- \frac{\xi_{\vec{k}}^2-g_{\vec{k}}^2-|\epsilon_{\vec{k}}|^2}{E_+ E_-} D_{++}] \bigg\{ \delta(\omega-2 E_+) + \delta(\omega + 2 E_+) \bigg\} \nonumber \\
& +   [C_{--}- \frac{\xi_{\vec{k}}^2-g_{\vec{k}}^2-|\epsilon_{\vec{k}}|^2}{E_+ E_-} D_{--}] \bigg\{ \delta(\omega-2 E_-) + \delta(\omega + 2 E_-) \bigg\} \nonumber \\
& + [C_{+-}- \frac{\xi_{\vec{k}}^2-g_{\vec{k}}^2-|\epsilon_{\vec{k}}|^2}{E_+ E_-} D_{+-}] \bigg\{ \delta(\omega-(E_+ + E_-))
+ \delta(\omega + (E_+ + E_-)) \bigg\},  \\
\mathcal{A}_2(\vec{k},\omega) & \equiv  B_{+-} \; \bigg\{ \delta(\omega-( E_+ + E_-)) + \delta(\omega + ( E_+  + E_-)) \bigg\}.
\end{align}
\end{subequations}
\end{widetext}
The $\vec{k}$ summation in Eq.~\eqref{eq:sigmaH1} is calculated numerically for each $\omega$ and the
results are plotted in Fig.~\ref{fig:ExpIm} over two different ranges of $\omega/t$ so that the details at larger $\omega/t$, where $|\mathrm{Im}\, \sigma_H^{(1)}|$ is smaller, can be clearly seen. $\mathrm{Im}\,\sigma_H^{(1)}(\omega)$
has several sign changes as a function of $\omega$ because  the different factors in Eq.~\eqref{eq:sigmaH1} change
sign at different $\vec{k}$ positions with different quasiparticle energies.  Also note that $\mathrm{Im}\,\sigma_H^{(1)}(\omega)$ is non-zero for arbitrarily small $\omega$ since the external field can excite quasiparticle pairs at arbitrarily small energy near the line nodes in the superconducting gap.
Although $\sigma_H^{(1)}(\omega)$ vanishes as $\omega \rightarrow 0$, 
this feature is not visible in Fig.~\ref{fig:ExpIm} (left panel) because the crossover to small $\omega$ behavior occurs at very small
frequency, $\omega < 0.01t$ (see Fig.6 of Ref.~\onlinecite{Yanase2016}).

\begin{figure*}[tp]
\centering
\begin{minipage}{0.5\textwidth}
\centering
\includegraphics[width=0.7\linewidth]{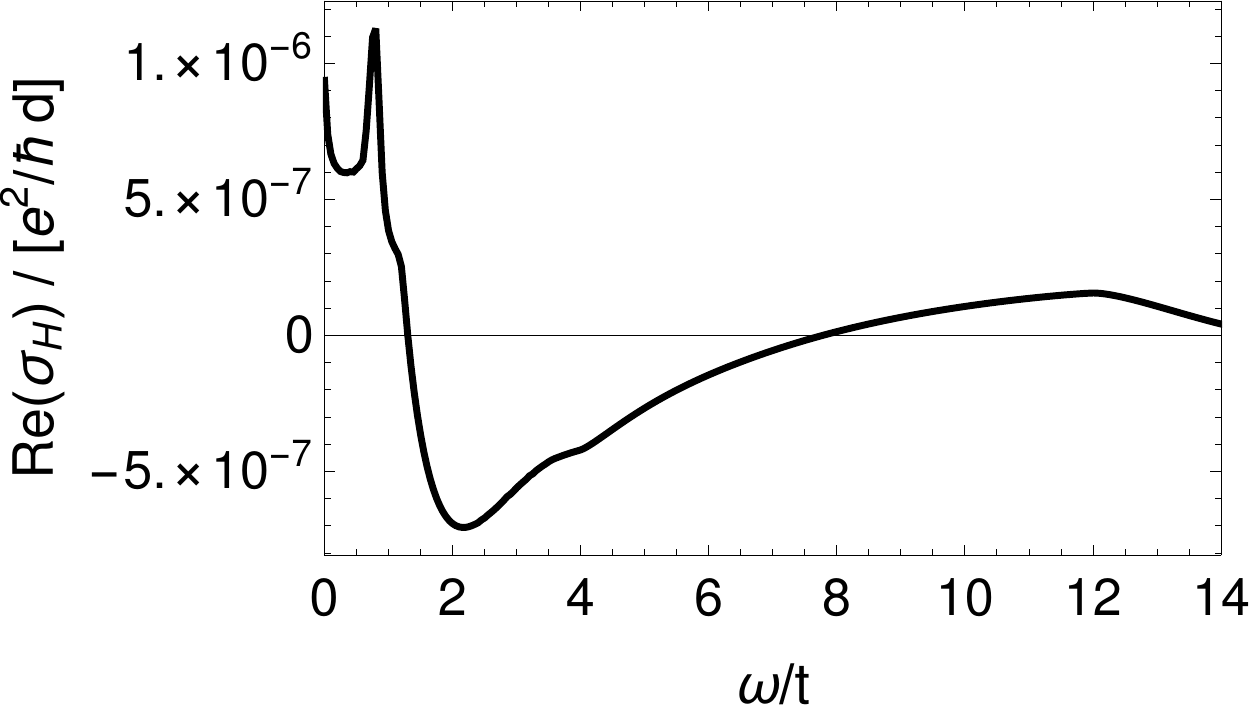}
\end{minipage}%
~
\begin{minipage}{0.5\textwidth}
\centering
\includegraphics[width=0.7\linewidth]{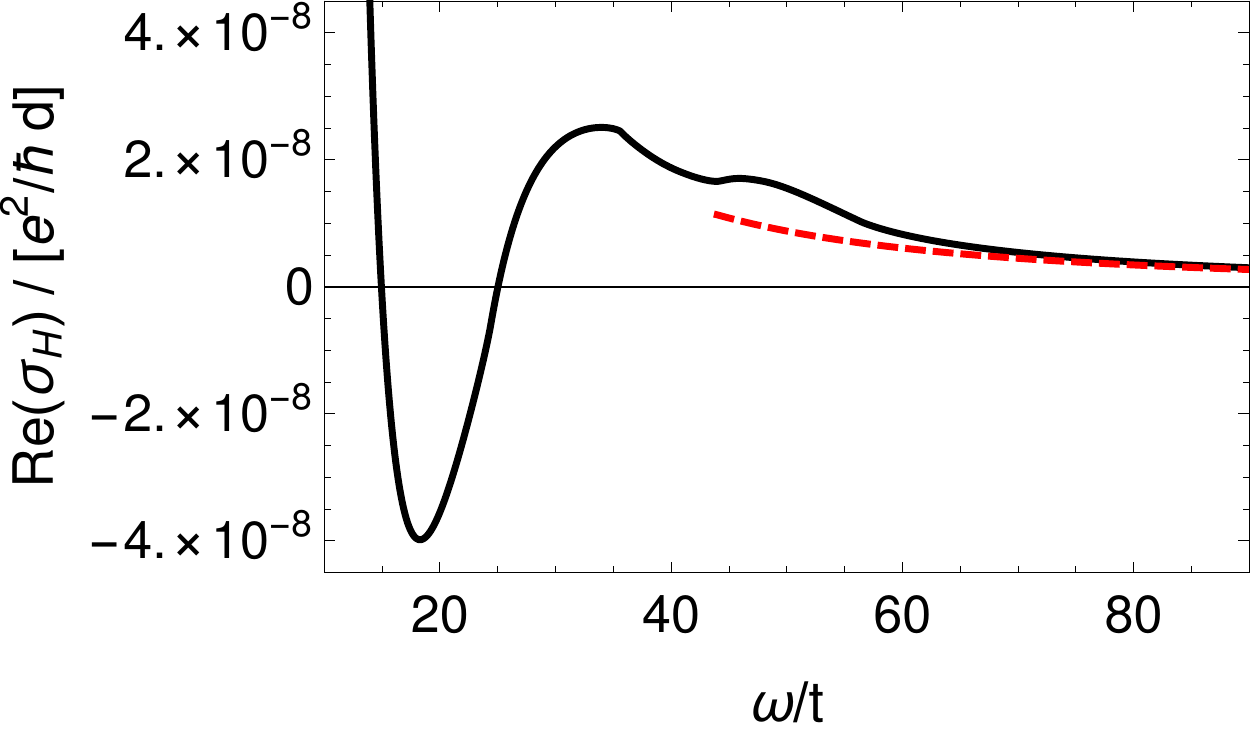}
\label{fig:ExpReL}
\end{minipage}
\caption{\label{fig:ExpRe}Numerical results for $\mathrm{Re} \; \sigma_H^{(1)}(\omega)$. Left panel: small frequency regime; right panel: large frequency regime.
Note that the scales of the vertical axis in the two figures are different. In the right figure the red dashed line is a high frequency asymptotic result.
Parameters used are the same as in Fig.~\ref{fig:ExpIm}.}
\end{figure*}

The real part, $\mathrm{Re} \,\sigma_H^{(1)}(\omega)$, can be computed from the data for $\mathrm{Im}\,\sigma_H^{(1)}(\omega)$ by the Kramers-Kronig transformation,
\begin{align}
\mathrm{Re}\,\sigma_H^{(1)}(\omega) & = \frac{2}{\pi} \mathcal{P} \int_0^\infty \frac{\nu \; \mathrm{Im}\;\sigma_H^{(1)}(\nu)}{\nu^2-\omega^2} d\nu,  \label{eq:KK1}
\end{align}
where $\mathcal{P}$ stands for Cauchy principal value integral. The results for $\mathrm{Re}\, \sigma_H^{(1)}(\omega)$ are plotted in Fig.~\ref{fig:ExpRe}.
In the right panel of Fig.~\ref{fig:ExpRe}, the red dashed line is an exact high frequency asymptotic result, whose expression is given by~\cite{Shastry1993}
\begin{gather}
\sigma_H(\omega \rightarrow \infty )=\frac{i }{ \omega^2} \langle [\hat{J}_x,\hat{J}_y] \rangle +\mathcal{O}(\frac{1}{\omega^4}), \label{eq:asymptotic1}
\end{gather}
where $[\hat{J}_x,\hat{J}_y]$ is an equal time commutator and the expectation value $\langle \cdots\rangle$ is with respect to the ground state of the BdG Hamiltonian.
In App.~\ref{sec:sumrule}, we compute $\langle [\hat{J}_x,\hat{J}_y] \rangle$ to first order in $\hat{\mathcal{H}}^\prime$ and find $\langle [\hat{J}_x,\hat{J}_y]\rangle^{(1)} \approx - i\, 2.2 \times 10^{-5} t^2 \, e^2/(\hbar \, d)$.
Similar to $\mathrm{Im}\,\sigma_H^{(1)}(\omega)$, $\mathrm{Re}\,\sigma_H^{(1)}(\omega)$ has further structure at very low frequency, $\omega < 0.01t$.
It saturates to a constant with a zero slope as $\omega \rightarrow 0$. Again, due to the large frequency range in Fig.~\ref{fig:ExpRe} (left panel),
this feature is not visible.
\subsection{Discussions of $\sigma_H^{(1)}$ }~\label{sec:dissigmaH1}

From Eq.~\eqref{eq:sigmaH1-2}, we can identify the necessary ingredients  for $\sigma_H^{(1)}$ to be nonzero. As emphasized previously, both the chiral $f$-wave and the chiral $d$-wave components need to be present.  In particular,  the dependence of $\sigma_H^{(1)}$ on these two parameters is through the combination
$i[f_{\vec{k}} d^*_{\vec{k}} -f^*_{\vec{k}}d_{\vec{k}}]$, which is proportional to the chirality.  Under time reversal, this combination,
and consequently $\sigma_H^{(1)}$, changes sign.  This can be seen explicitly from the fact that under time reversal,
$\Delta_{12}(\vec{k}) \rightarrow  - \Delta_{12}^*(-\vec{k})$, $\Delta_{21}(\vec{k}) \rightarrow  - \Delta_{21}^*(-\vec{k})$
and $2i[f_{\vec{k}} d^*_{\vec{k}} -f^*_{\vec{k}}d_{\vec{k}}]=\Delta_{21}(\vec{k})\Delta_{21}^*(\vec{k})-\Delta_{12}(\vec{k})\Delta_{12}^*(\vec{k})$.  This is the only combination quadratic in $\Delta_{12}$ and/or $\Delta_{21}$ that is odd under time-reversal.  It is also this term that makes the order parameter $\hat{\Delta}(\vec{k})$ nonunitary.

The second important ingredient for $\sigma_H$  is the complex inter-sublattice hopping, $\epsilon_{\vec{k}}$, since both velocity terms appearing in Eq.~\eqref{eq:sigmaH1-2},
$\vec{h}\cdot \partial_{k_x} \vec{h}\times \partial_{k_y} \vec{h}$ and $\Omega_{xy}$, vanish if $\epsilon_{\vec{k}}$ is real.  These velocity terms are consistent with another general requirement for $\sigma_H$ to be nonzero in the multi-band mechanism.  Namely, some antisymmetrized products of the velocity operators, $v^x_{ab} v^y_{cd} - v^y_{ab} v^x_{cd}$ (where $a,b$ label orbitals or, in our case, sublattices) need to be nonzero.
Note that SOC, $g_{\vec{k}}$, is not necessary for a nonzero $\sigma_H$.
Of the two terms in Eq.~\eqref{eq:sigmaH1-2}, only the first term vanishes if $g_{\vec{k}}=0$.
The second term, with $\Omega_{xy}$, only depends on $g_{\vec{k}}$ through the Bogoliubov quasiparticle energies $E_{\pm}$ and remains nonzero if the SOC is absent.

The two key ingredients identified above, the mixing of the chiral $f$- and $d$-wave order parameters and the complex
inter-sublattice hopping, $\epsilon_{\vec{k}}$, are both direct consequences of the non-symmorphic symmetry of $\mathrm{UPt_3}$.
They would both be absent if the lattice were symmorphic. In this sense, the terms that we have identified for $\sigma_H$
are unique to non-symmorphic chiral superconductors.

\begin{figure*}[htp]
\centering
\includegraphics[width=0.65\textwidth]{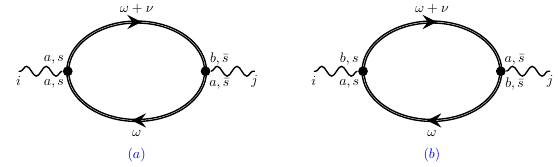}
\caption{Diagramatic representation of the non-vanishing contributions to $\sigma_H$,  where wiggly lines are photons and double solid lines with arrows are Green's
functions given by Eq.~\ref{eq:Ginv1}.   The photon polarization is labelled by $i,j=x,y$. $a,b$ are sublattice labels and $s$ is the spin label.  If $s=\{\uparrow,\downarrow \}$, then $\bar{s}=\{\downarrow,\uparrow \}$.
Note that, in each diagram, the spin labels on a right vertex are opposite to that on the corresponding left vertex.
This is because, in each diagram, each Green's function contributes one superconducting order parameter that pairs electrons of opposite spin,
while all normal state Hamiltonian matrix elements, including SOC, only connect electrons of the same spin.
}
\label{fig:Bubble1}
\end{figure*}

The two terms in Eq.~\eqref{eq:sigmaH1-2} can be represented by Feynman diagrams, which are shown in Fig.~\ref{fig:Bubble1}.
For each diagram in Fig.~\ref{fig:Bubble1}, the time-reversed diagram needs to be subtracted.
There are two types of diagrams. In Fig.~\ref{fig:Bubble1}(a) only
one of the two vertices involves two different orbitals; while in Fig.~\ref{fig:Bubble1}(b) both the  vertices
involve transitions between different orbitals.
Of the two terms in Eq.~\eqref{eq:sigmaH1-2}, the term $\propto \Omega_{xy}$ only contributes to Fig.~\ref{fig:Bubble1}(b), while the other term, $\propto \vec{h}\cdot \partial_{k_x} \vec{h}\times \partial_{k_y} \vec{h}$, is a mixture
of  Fig.~\ref{fig:Bubble1}(a) and ~\ref{fig:Bubble1}(b). This is because $\vec{h}\cdot
\partial_{k_x} \vec{h}\times \partial_{k_y} \vec{h}$ can be written
as a sum of $\epsilon_{\vec{k}}  \partial_{k_x} \epsilon^*_{\vec{k}} \partial_{k_y} g_{\vec{k}} +
\epsilon^*_{\vec{k}}  \partial_{k_x} g_{\vec{k}} \partial_{k_y} \epsilon_{\vec{k}} -\big\{ x \leftrightarrow y \big\}$
and
$g_{\vec{k}}  \partial_{k_x} \epsilon_{\vec{k}} \partial_{k_y} \epsilon^*_{\vec{k}} -\big\{ x \leftrightarrow y \big\}$,
of which the former and latter correspond to Fig.~\ref{fig:Bubble1}(a) and Fig.~\ref{fig:Bubble1}(b), respectively.
In the band basis, the $\Omega_{xy}$ term in Eq.~\eqref{eq:sigmaH1-2} corresponds to Fig.~\ref{fig:Bubble1}(a) (with $i,j$ now labelling bands), rather than Fig.~\ref{fig:Bubble1}(b) as in the orbital basis;
while the whole $\vec{h}\cdot\partial_{k_x}\vec{h}\times \partial_{k_y}\vec{h}$ term corresponds to Fig.~\ref{fig:Bubble1}(b).  It is clear in the band basis that both Fig.~\ref{fig:Bubble1}(a) and Fig.~\ref{fig:Bubble1}(b)
vanish if the inter-band pairing is zero, similar to what was found in Ref.~\onlinecite{Taylor2012}.

Note that Fig.~\ref{fig:Bubble1}(b) type of diagram is absent in Ref.~\onlinecite{Taylor2012} because the model studied there has a real inter-orbital hopping $\epsilon_{\vec{k}}$, which makes the contribution from Fig.~\ref{fig:Bubble1}(b)
with the photon polarization $(i,j)=(x,y)$ exactly cancel the same diagram with $(i,j)=(y,x)$.  On the other hand,
Fig.~\ref{fig:Bubble1}(a) vanishes in the current model unless $\epsilon_{\vec{k}}$ is complex, while it survives in Ref.~\onlinecite{Taylor2012} for real inter-orbital hopping, due to the different way the inter-orbital pairing arises in the two models.

$\sigma_H(\omega)$ also needs to obey the following two sum rules~\cite{Lange1999,Drew1997},
\begin{align}
\int_{0}^\infty d\omega \; \mathrm{Re} \; \sigma_H(\omega) & =0, \label{eq:sumRe} \\
\int_{-\infty}^{\infty} d\omega \; \frac{\omega \mathrm{Im} \; \sigma_H(\omega)}{\pi } &= - i \langle [\hat{J}_x,\hat{J}_y]\rangle.\label{eq:sumIm}
\end{align}
where  Eq.~\eqref{eq:sumIm} is analogous to the well-known optical
conductivity $f$-sum rule. In App.~\ref{sec:sumrule}, we show these sum rules are satisfied, both analytically and numerically, by $\sigma_H^{(1)}(\omega)$.

Lastly we mention that the Hall conductivity, quite generally, needs to satisfy several symmetry constraints.
Under time reversal, all vertical
mirror reflections, and particle-hole interchange, $\sigma_H$ must reverse its sign. Both $\sigma_H^{(1)}$ given in Eq.~\eqref{eq:sigmaH1-2},
and the full Green's function result of $\sigma_H$ given in App.~\ref{sec:appendixFullG} are consistent with these symmetry constraints.

\section{Estimation of the Kerr rotation angle $\theta_K$}~\label{sec:Kerrangle}

From the numerical results of $\sigma_H(\omega)$, the Kerr rotation angle, $\theta_K$, can be estimated using Eq.~\eqref{eq:Kerr}, which also involves the complex index of refraction, $n(\omega)$.
Here we use our results to estimate the Kerr angle for $\mathrm{UPt_3}$, where
$\theta_K$ was measured~\cite{Schemm2014} at a laser frequency
 $\omega \approx 0.8 \; \mathrm{eV}$.

We first estimate $n(\omega$=$0.8\mathrm{eV})$ from experimental data. By definition $n(\omega) =\sqrt{\epsilon(\omega)}$, where $\epsilon(\omega)$ is related to the conductivity, $\sigma(\omega)$, by $\epsilon(\omega) =\epsilon_{\infty}+i 4\pi \sigma(\omega)/\omega$ and $\epsilon_{\infty}$ is the high frequency limit
dielectric constant. We extract $\sigma(\omega$=$0.8\mathrm{eV})  \approx (1.7 +i \, 0.4)\times 10^{15} \; \mathrm{s^{-1}}$ from the experimental data of Ref.~\onlinecite{Sulewski1988}.
Taking $\epsilon_{\infty}=1$, we obtain  $\epsilon(\omega$=$0.8 \mathrm{eV}) \approx -3.1+i \, 17.5$, which gives an index of refraction,
\begin{gather}
n(\omega=0.8 eV) \approx 2.7 + i\, 3.2.  \label{eq:indexn}
\end{gather}

To obtain a value for $\sigma_H(\omega\approx 0.8 \mathrm{eV})$, we need to estimate the in-plane hopping parameter $t$ in eV, since we have scaled all energies by $t$.
This can be obtained by comparing the normal state band dispersions of our two-band model along the symmetry directions $\mathrm{A-L-H-A}$ in the $k_z=\pi$ plane
to the corresponding first-principle calculation results from Ref.~\onlinecite{Nomoto2016}. The comparison gives $t\approx 36 \mathrm{meV}$ (for
details, see App.~\ref{sec:testimation}).
This value of $t$ corresponds to $\omega/t \approx 22.2$ at $\omega=0.8\mathrm{eV}$. From our numerical results for $\sigma_H(\omega)$ in Fig.~\ref{fig:ExpIm} and
Fig.~\ref{fig:ExpRe} we obtain, at $\omega/t\approx 22.2$ ,
\begin{gather}
\sigma_H(\omega \approx 0.8 \mathrm{eV}) \approx -(2.3+i\, 5.1)\times 10^{-8} \frac{e^2}{\hbar\,d}, \label{eq:sigmaH08eV}
\end{gather}
where $d=4.9\AA$ is the $c$-axis lattice spacing of $\mathrm{UPt_3}$.
From Eqs.~\eqref{eq:indexn},~\eqref{eq:sigmaH08eV} and ~\eqref{eq:Kerr}, the Kerr angle is then,
\begin{align}
\theta_K  & \approx  34 \times 10^{-9} \, \mathrm{rad}.
\end{align}

Our estimated $\theta_K$ is about an order of magnitude smaller than the experimental value of about 350 nanoradians measured at the lowest temperatures~\cite{Schemm2014}.
 However, it may still be a significant contribution to the explanation for the Kerr measurement on $\mathrm{UPt_3}$~\cite{Schemm2014}
given that there are uncertainties in the optical constants, the band parameters, and the magnitude of $\Delta_0$ used for this estimate. We briefly comment on these uncertainties.

Ideally, one would like measurements of $n(\omega)$ on the same crystal used for the Kerr measurements. Other optical data on $\mathrm{UPt_3}$ would give somewhat different
results~\cite{Schoenes1985,Dressel2002,Marabelli1986}, although we estimate that the uncertainty in the optical data is unlikely to change the estimated Kerr angle by more than a factor of 3 or so.

As to the band parameters,  uncertainty comes both from the value of $t$ and from the fact that a very simplified nearest-neighbour hopping model has been used to approximate the two bands which give rise to the starfish Fermi surface. This likely introduces a larger uncertainty than that from errors in the estimate of $n(\omega)$.

The other parameter that can greatly affect the size of $\theta_K$ is $\Delta_0$, the amplitude of the gap function written in the orbital basis.  Note that $\Delta_0$  is not the gap that one would observe in tunneling measurements.  Defining $\Delta_g$ as the position of the coherence peak in the Bogoliubov quasiparticle density of states spectrum, one finds $\Delta_g \approx 0.16\, \Delta_0$
(see Fig.6 of Ref.~\onlinecite{Yanase2016}).   Experiments have found values for $\Delta_g$ of  0.04 meV~\cite{Goll1993}, 0.1 meV~\cite{Goll1995}, and more recently, 0.5 meV~\cite{Gouchi2015}.  The parameters we used, taken from Yanase~\cite{Yanase2016}, with $t=36$ meV, corresponds to $\Delta_g=0.58$ meV, roughly consistent with the most recent experimental value.
Since the Kerr angle scales quadratically with the gap magnitude, smaller values of $\Delta_g$ would give much smaller values of $\theta_K$.
For example, setting $T_{\rm c}=0.53$K, we find $\Delta_g\approx 0.11$ meV for our model in the weak-coupling limit, which would reduce $\theta_K$ by a factor of 26.

Lastly there are several other Fermi surface sheets that we did not take into account, which might contribute to $\theta_K$.  These additional contributions could either increase or decrease the total $\theta_K$, depending on their relative magnitude and sign.

With these uncertainties in mind, we conclude that the $\theta_K$ that we have identified here can be significant for explaining the Kerr measurement on $\mathrm{UPt_3}$, even if it is not large enough to account for the whole experimentally observed signal.  Further experiments
and theoretical studies are needed to resolve the above uncertainties.

\section{Conclusion and Discussions}~\label{sec:conclusion}

To summarize, by considering a simplified two band model that results from ABAB stacking for the starfish-like Fermi surface of $\mathrm{UPt_3}$,
we have identified a contribution to the ac anomalous Hall conductivity for $\mathrm{UPt_3}$ within the intrinsic multiband chiral superconductivity mechanism.
The Kerr angle estimated from the computed Hall conductivity
can be significant for understanding the Kerr measurement on $\mathrm{UPt_3}$.
This mechanism requires non-zero interband pairing. Since intra- and inter-band pairing are indistinguishable at the six points on the $k_z=\pm \pi$ plane where the starfish-like Fermi surfaces of UPt$_3$ intersect, this is a useful model for studying the multiband chiral superconductivity mechanism.

We have identified two crucial ingredients for the nonzero $\sigma_H$: a complex inter-sublattice hopping between  $\mathrm{U}$ sites and a novel superconducting
order parameter that involves mixing between chiral $f$-wave and chiral $d$-wave pairing. Both of these are consequences of the nonsymmorphic group symmetry of
the $\mathrm{UPt_3}$ crystal lattice. If the inter-sublattice hopping is real or if one of the chiral $f$- and $d$-wave pairing components is absent, then
$\sigma_H$ and $\theta_K$ vanish. This is a generalization of, albeit somewhat distinct from, the multiband chiral superconductivity mechanism for the anomalous ac Hall effect
in a chiral $p$-wave superconductor~\cite{Taylor2012}. The $\sigma_H$ and $\theta_K$ contribution that we have discussed here can also be applied to other nonsymmorphic chiral superconductors.

In our analysis we have identified two types of terms that contribute to $\sigma_H(\omega)$ at each $\vec{k}$ point, as can be seen from Eq.~\eqref{eq:sigmaH1-2}.
One term  does not require SOC, while the other does. The two make comparable contributions to $\sigma_H$. However, these two contributions in general can have
different signs at different $\vec{k}$ points, which results in multiple sign changes of $\sigma_H(\omega)$ as a function of $\omega$.
Because of these sign changes
the estimated Kerr angle can be sensitive to the band parameters as well as to the laser frequency used in the Kerr measurement. Therefore future Kerr measurements
at different frequencies would be very helpful in determining how relevant the Kerr angle contribution identified here is to $\mathrm{UPt_3}$.

We should mention that in our calculation we have neglected a small chiral $p$-wave component pairing in the original proposed superconducting order parameter
of Ref.~\onlinecite{Yanase2016}. This component is also symmetry allowed but is expected to be energetically less favorable compared with the dominant chiral $f$- and $d$-components.
In the two band model we consider, this $p$-wave component alone can also give rise to a nonzero $\sigma_H(\omega)$. This contribution relies on the nonunitary nature of the $p$-wave pairing (it pairs only one spin component if $\eta=1$), and requires nonzero SOC and complex inter-sublattice hopping. Presumably the admixture of this neglected small $p$-wave component will not significantly alter the estimated Kerr angle simply because its pairing amplitude is thought to be very small..

Recently the authors of Ref.~\onlinecite{Joynt2017} suggested that the Kerr rotation in $\mathrm{UPt_3}$ can not be understood without invoking pairing in completely filled or empty bands because the laser frequency used in the Kerr angle measurement~\cite{Schemm2014}, $\omega \approx 0.8 \,\mathrm{eV}$,
is bigger than the normal state bandwidth of the partially filled bands of $\mathrm{UPt_3}$. However,
this does not need to be the case for two reasons. First, since the incident photon breaks a Cooper pair and generates two Bogoliubov quasiparticles, the maximum energy cost is not the bandwidth, but twice the energy difference between the Fermi level and the bottom or top of the band (whichever is greater).  From Ref.~\onlinecite{Wang1987}, this maximum energy along the symmetry direction $\mathrm{A-L-H-A}$ in the $k_z=\pi$ plane is about $ 0.68 \,\mathrm{eV}$,
while from Ref.~\onlinecite{Nomoto2016}, this is about $0.84 \, \mathrm{eV}$.   The latter (which we used to determine the hopping $t$ in our model) allows energy-conserving transitions within the band at 0.8 eV.  Second, both $\mathrm{Re}\,\sigma_H(\omega)$ and $\mathrm{Im}\,\sigma_H(\omega)$ can
make significant contributions to $\theta_K$. Even if the laser frequency is larger than the excitation energy of two quasiparticles within the band, $\mathrm{Re}\,\sigma_H(\omega)$ will still
be nonzero at $\omega=0.8\,\mathrm{eV}$. Consequently, the observation of nonzero
$\theta_K$ in $\mathrm{UPt_3}$ at 0.8 eV may still be understood within a model of partially filled bands.

\section{Acknowledgements}
We would like to thank Tom Timusk and Steve Kivelson for helpful discussions.
This work is supported in part by NSERC (CK and ZW), the Canada Research Chair program (CK), the National Science Foundation under Grant 
No. NSF PHY11-25915 (AJB, CK, GZ), the Gordon and Betty Moore Foundation's EPiQS Initiative through Grant GBMF4302 (AJB and CK), the ANR-DFG grant Fermi-NESt  (GZ),  
and a grant from the Simons Foundation ($\#$395604 to CK). AJB, CK and GZ greatly appreciate the hospitality provided by the Kavli Institute for Theoretical Physics 
at UCSB and (for AJB, CK and ZW) the hospitality of the Stanford Institute for Theoretical Physics, where part of the work was completed.

\appendix \label{sec:appendix}
\section{Full Green's function calculation of $\sigma_H(\omega)$} \label{sec:appendixFullG}

As mentioned in the main text, the full Green's function calculation is much more involved than the perturbative calculation.
Here, we present some main steps for the full calculation of $\sigma_H(\omega)$, omitting detail of derivations.

We first establish some notation. We denote the four Bogoliubov quasiparticle energies of the BdG Hamiltonian
$\hat{\mathcal{H}}^{(a)}(\vec{k})$, from Eq.~\eqref{eq:Hblock1} of the main text, as $E_i$, with $i=\{1,2,3,4\}$.
The $E_i$ are solutions to
\begin{gather}
\mathrm{det}\big\{ \omega - \hat{\mathcal{H}}^{(a)}(\vec{k})\big\}=0,
\end{gather}
which can be expanded as
\begin{gather}
\omega^4+\alpha_{\vec{k}} \; \omega^2+\beta_{\vec{k}} \; \omega +\gamma_{\vec{k}}=0.  \label{eq:quartic}
\end{gather}
where the three coefficients are given by
\begin{widetext}
\begin{subequations}
\begin{align}
\alpha_{\vec{k}} & = -2 (\xi_{\vec{k}}^2 +g_{\vec{k}}^2+|f_{\vec{k}}|^2 +|d_{\vec{k}}|^2+|\epsilon_{\vec{k}}|^2 ), \\
\beta_{\vec{k}}  & = 4 i\, (f_{\vec{k}} d_{\vec{k}}^* -f_{\vec{k}}^* d_{\vec{k}})\, g_{\vec{k}}, \label{eq:betak} \\
\gamma_{\vec{k}}  & =  (\xi_{\vec{k}}^2 - g_{\vec{k}}^2 -|\epsilon_{\vec{k}}|^2 +|f_{\vec{k}}|^2+|d_{\vec{k}}|^2)^2 +  4 |d_{\vec{k}}|^2 |\epsilon_{\vec{k}}|^2 \nonumber \\
 & +(|f_{\vec{k}}|^2 -|d_{\vec{k}}|^2)(\epsilon_{\vec{k}} +\epsilon_{\vec{k}}^*)^2 + (f_{\vec{k}}^* d_{\vec{k}} - f_{\vec{k}} d_{\vec{k}}^*)^2
 -i(f_{\vec{k}} d_{\vec{k}}^* +f_{\vec{k}}^* d_{\vec{k}})\,
(\epsilon_{\vec{k}}^2- (\epsilon_{\vec{k}}^*)^2) .
\end{align}
\end{subequations}
\end{widetext}
Eq.~\eqref{eq:quartic} is a quartic equation for $\omega$ rather than a quadratic equation in $\omega^2$ due to the $\beta_{\vec{k}} \omega$ term.
Because of this, the solutions $E_i$ do not occur as $\{+E,-E\}$ particle hole pairs.
However, this does not contradict the particle-hole symmetry of the full superconducting BdG Hamiltonian
which is restored when $\hat{\mathcal{H}}^{(a)}$ is combined with the other $4\times 4$ block $\hat{\mathcal{H}}^{(b)}(\vec{k})$,
given in Eq.~\eqref{eq:Hblock2}, to form the full $\hat{\mathcal{H}}_{\mathrm{BdG}}$.
Also, because of the $\beta_{\vec{k}}\omega$ term in Eq.~\eqref{eq:quartic}, the expressions for the $E_i$, in terms of the three
coefficients $\{\alpha_{\vec{k}},\beta_{\vec{k}},\gamma_{\vec{k}}\}$ are much more complicated than in the case of $\beta_{\vec{k}}=0$. For brevity we will not present
them here.

With the coefficients $\{\alpha_{\vec{k}},\beta_{\vec{k}},\gamma_{\vec{k}}\}$ and $E_i$ defined above we can now write the final result for
$\sigma_H(\omega)$ as follows
\begin{widetext}
\begin{align}
\sigma_H(\omega) & = \sum_{\vec{k}} 16 i \;  \xi_{\vec{k}} \,\vec{h}\cdot\partial_{k_x}\vec{h}\times \partial_{k_y}\vec{h}
  \bigg\{ \widetilde{F}_1(\vec{k},\omega)+(\xi_{\vec{k}}^2-g_{\vec{k}}^2-|\epsilon_{\vec{k}}|^2)
\widetilde{F}_2(\vec{k},\omega) \bigg\}  \nonumber \\
 & + 4 i \; \xi_{\vec{k}} \, (f_{\vec{k}} d_{\vec{k}}^* -f_{\vec{k}}^* d_{\vec{k}}) \,  \Omega_{xy} \, \widetilde{F}_3(\vec{k},\omega)
 - 8 \xi_{\vec{k}} \; \mathcal{O}_h(\vec{k}) \; \widetilde{F}_2(\vec{k},\omega),   \label{eq:appendixSigmaH}
\end{align}
\end{widetext}
where $\Omega_{xy}$ was defined in Eq.~\eqref{eq:Oxy}. In Eq.~\eqref{eq:appendixSigmaH} the three frequency dependent functions
are defined as
\begin{widetext}
\begin{subequations}
\begin{align}
\widetilde{F}_1(\vec{k},\omega) & = - \frac{1}{2}\sum_{i=1}^4 |E_i| \frac{\omega^4 -\omega^2(4E_i^2-\alpha_{\vec{k}})+(3 E_i^4 -\alpha_{\vec{k}} E_i^2 +3\gamma_{\vec{k}})}
{\prod_{j=1,j\ne i}^4 (E_j - E_i)\big\{ (E_j-E_i)^2 -\omega^2 \big\}}, \\
\widetilde{F}_2(\vec{k},\omega) & = -\frac{1}{2} \sum_{i=1}^4 |E_i| \frac{-2\omega^2 + (9 E_i^2 +\alpha_{\vec{k}} +\gamma_{\vec{k}}/E_i^2)}
{\prod_{j=1,j\ne i}^4 (E_j - E_i)\big\{ (E_j-E_i)^2 -\omega^2 \big\}}, \\
\widetilde{F}_3(\vec{k},\omega) & = - \frac{1}{2} \sum_{i=1}^4 \mathrm{sgn}(E_i) \frac{\omega^4 -\omega^2(6 E_i^2-\alpha_{\vec{k}})+(12 E_i^4+4 \gamma_{\vec{k}})}
{\prod_{j=1,j\ne i}^4 (E_j - E_i)\big\{ (E_j-E_i)^2 -\omega^2 \big\}}. \label{eq:tildeFi}
\end{align}
\end{subequations}
\end{widetext}
$\widetilde{F}_1, \widetilde{F}_2$ and $\widetilde{F}_3$ are connected to the three functions, $F_i(\vec{k},\omega)$, that we introduced in our
perturbative calculations, by
\begin{widetext}
\begin{align}
\widetilde{F}_1(\vec{k},\omega) & =\frac{\beta_{\vec{k}}}{2} \, F_1(\vec{k},\omega)  +\mathcal{O}( \beta_{\vec{k}}^3) \; , \; 
\widetilde{F}_2(\vec{k},\omega)  = - \frac{\beta_{\vec{k}}}{2 \, E_+  E_-} \, F_2(\vec{k},\omega) +\mathcal{O}( \beta_{\vec{k}}^3) \; , \; 
\widetilde{F}_3(\vec{k},\omega)  =  F_3(\vec{k},\omega) +\mathcal{O}( \beta_{\vec{k}}^2), \label{eq:F3Exp}
\end{align}
\end{widetext}
where  $E_{\pm}$ are the two Bogoliubov quasiparticle energies of the zeroth order Hamiltonian(see Eq.~\eqref{eq:Epm}).
From these relations we see that the parameter that controls our perturbative calculation is $\beta_{\vec{k}}$ rather than simply $d_{\vec{k}}$.

The $\mathcal{O}_h(\vec{k})\widetilde{F}_2(\vec{k},\omega)$ term in Eq.~\eqref{eq:appendixSigmaH}
contains terms of higher powers, fourth order in $f_{\vec{k}}$ and $d_{\vec{k}}$, compared with the other terms
that are second order in $f_{\vec{k}}$ and $d_{\vec{k}}$ (ignoring the $f_{\vec{k}}$ dependence through the quasiparticle energies $E_{\pm}$).
 This is clear from Eq.~\eqref{eq:betak}, the expression for $\beta_{\vec{k}}$, and from
\begin{align}
\hspace{-0.2cm}
\mathcal{O}_h(\vec{k}) & = (|f_{\vec{k}}|^2+|d_{\vec{k}}|^2) \, g_{\vec{k}} \,\Omega_{xy} \nonumber \\
& - (|f_{\vec{k}}|^2 -|d_{\vec{k}}|^2) \,
\big\{ \mathrm{Re}[\epsilon_{\vec{k}}] \Omega_{xy}^{(1)} + \mathrm{Im}[\epsilon_{\vec{k}}] \Omega_{xy}^{(2)}   \big\}  \nonumber \\
& + (f_{\vec{k}} d_{\vec{k}}^* +f_{\vec{k}}^* d_{\vec{k}}) \big\{ \mathrm{Re}[\epsilon_{\vec{k}}] \Omega_{xy}^{(2)}
-\mathrm{Im}[\epsilon_{\vec{k}}] \Omega_{xy}^{(1)}   \big\},
\end{align}
where we have introduced two additional anti-symmetrized velocity products $\Omega_{xy}^{(1)}$ and $\Omega_{xy}^{(2)}$, defined as follows
\begin{subequations}
\begin{align}
\Omega_{xy}^{(1)} & = 2 \left\{ \partial_{k_x} g_{\vec{k}} \; \partial_{k_y} \mathrm{Im}[\epsilon_{\vec{k}}] -\partial_{k_x} \mathrm{Im}[\epsilon_{\vec{k}}] \; \partial_{k_y} g_{\vec{k}}  \right\}, \\
\Omega_{xy}^{(2)} & = 2 \left\{ \partial_{k_x} g_{\vec{k}} \; \partial_{k_y} \mathrm{Re}[\epsilon_{\vec{k}}] - \partial_{k_x} \mathrm{Re}[\epsilon_{\vec{k}}] \;
\partial_{k_y} g_{\vec{k}} \right\}. \label{eq:Oxy2def}
\end{align}
\end{subequations}

\begin{figure*}[tp]
\centering
\begin{minipage}{0.5\textwidth}
\centering
\includegraphics[width=0.75\linewidth]{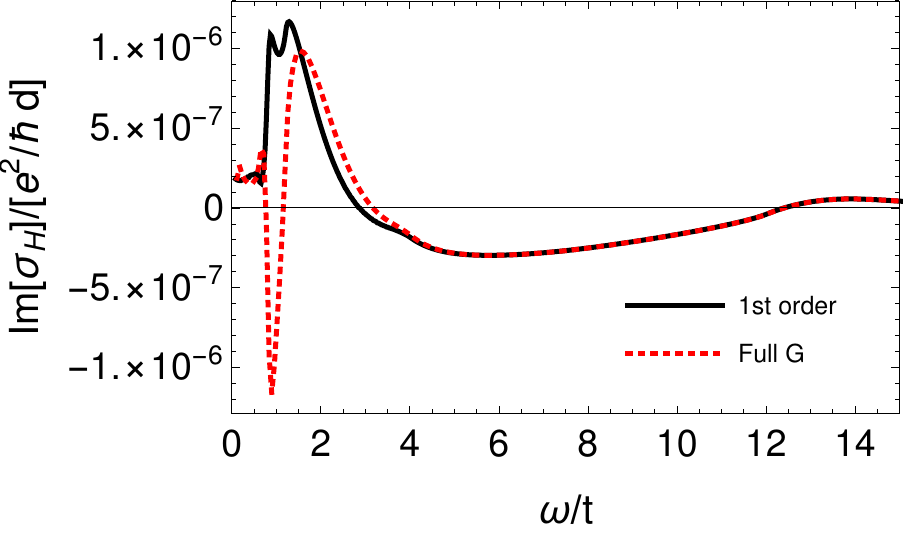}
\label{fig:compareImS}
\end{minipage}%
~
\begin{minipage}{0.5\textwidth}
\centering
\includegraphics[width=0.75\linewidth]{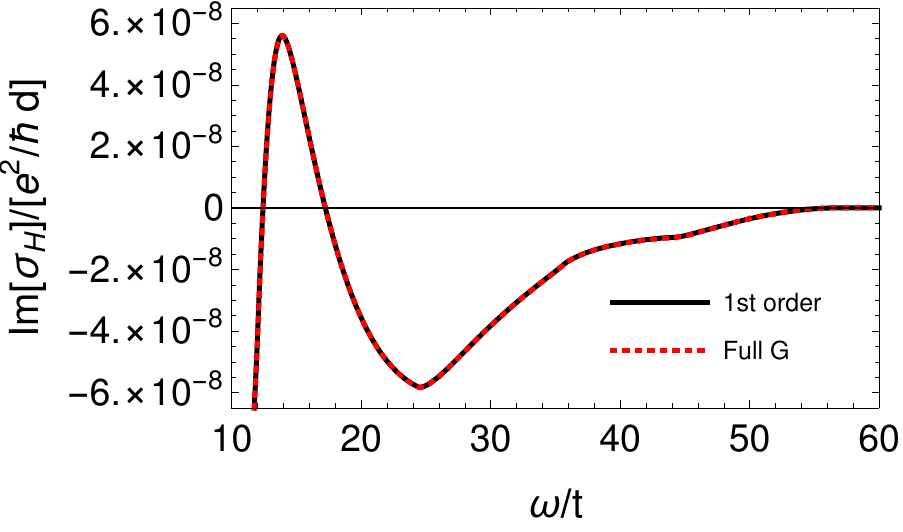}
\label{fig:compareImL}
\end{minipage}
\caption{\label{fig:compareIm}Comparison between the numerical results for $\mathrm{Im} \,\sigma_H^{(1)}$(thick black line) and that
for $\mathrm{Im} \, \sigma_H$(dashed red line). Left panel: small $\omega/t \le 14$; right panel: large $\omega/t \ge 10$. Notice that
the vertical axis scales of the two panels are different. Parameters used are the same as in Fig.~\ref{fig:ExpIm}. }
\end{figure*}

From $\sigma_H(\omega+i\delta)$ in Eq.~\eqref{eq:appendixSigmaH} we can derive its imaginary part, $\mathrm{Im} \, \sigma_H(\omega)$.
Then we can numerically evaluate $\mathrm{Im} \, \sigma_H(\omega)$ and compare the results with our perturbation results for $\mathrm{Im} \;\sigma_H^{(1)}(\omega)$ in the main text.
The comparison is given in Fig.~\ref{fig:compareIm}.  We see that the two are quite different for $\omega \lesssim 2\, t$, but they are essentially indistinguishable for $\omega\gtrsim 4 t$.

\begin{figure*}[tp]
\centering
\begin{minipage}{0.5\textwidth}
\centering
\includegraphics[width=0.75\linewidth]{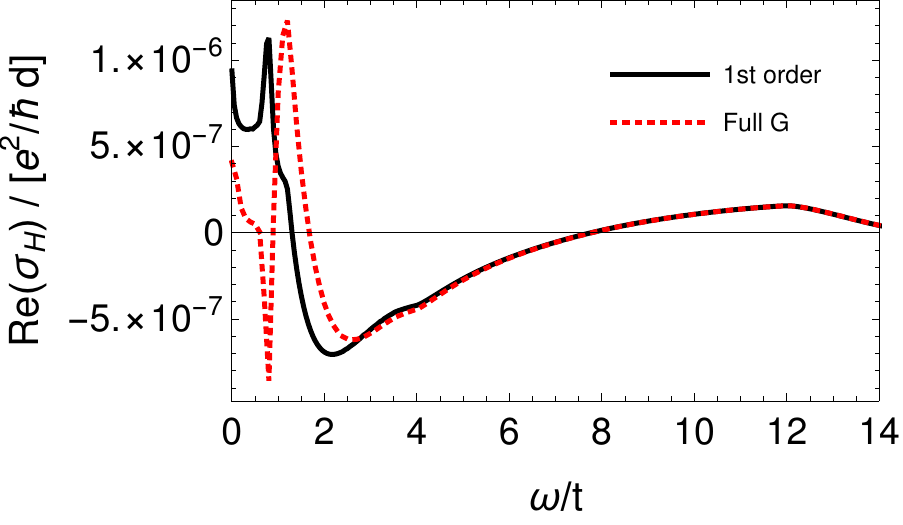}
\label{fig:compareReS}
\end{minipage}%
~
\begin{minipage}{0.5\textwidth}
\centering
\includegraphics[width=0.75\linewidth]{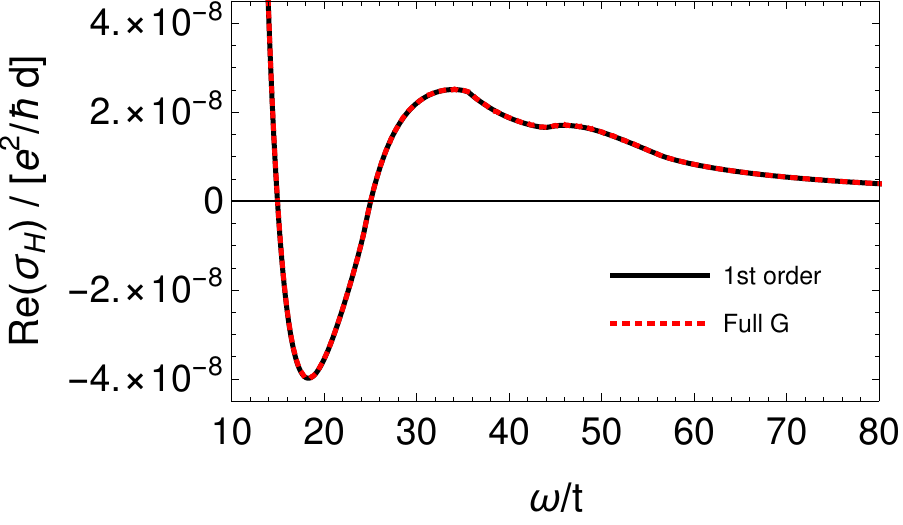}
\label{fig:compareReL}
\end{minipage}
\caption{\label{fig:compareRe}Comparison between the numerical results for $\mathrm{Re} \,\sigma_H^{(1)}$(thick black line) and that for
$\mathrm{Re} \, \sigma_H$(dashed red line). Left panel: small $\omega/t \le 14$; right panel: large $\omega/t \ge 10$. Note that
the vertical axis scales of the two panels are different.}
\end{figure*}

We can also compute $\mathrm{Re}\,\sigma_H(\omega)$ by the Kramers-Kronig transformation and compare
the results with $\mathrm{Re}\,\sigma_H^{(1)}$, presented in the main text. This comparison is shown in Fig.~\ref{fig:compareRe}. Again at $\omega \gtrsim 4t$ the two agree well.

\section{Derivation of $\sigma_H^{(1)}$}\label{sec:appendixPixy}

In order to compute $\sigma_H^{(1)}$, using Eq.~\eqref{eq:sigmadef} and~\eqref{eq:pixy1}, we introduce the function $\mathcal{F}_{xy}^{(1)}(\vec{k};i\omega_n, i\nu_m)$
such that
\begin{align}
 & \pi_{xy}^{(1)}(i\nu_m) -\pi_{yx}^{(1)}(i\nu_m)
=  T \sum_{\vec{k},\omega_n} \mathcal{F}_{xy}^{(1)}(\vec{k};i\omega_n, i\nu_m). \label{eq:pxympyx}
\end{align}
From the expression for $\pi_{xy}^{(1)}(i\nu_m)$ in Eq.~\eqref{eq:pixy1}, we can write $\mathcal{F}_{xy}^{(1)}$ as follows
\begin{widetext}
\begin{align}
\mathcal{F}_{xy}^{(1)} &  \equiv \big\{ \mathrm{Tr}[\hat{v}^x \hat{G}^{(0)}(\vec{k}, i\omega_n+i\nu_m) \hat{v}^y \hat{G}^{(1)}(\vec{k}, i\omega_n)]
 + \big\{ (0) \leftrightarrow (1) \big\} \big\} - \big\{ x \leftrightarrow y \big\}.
\label{eq:Fxy1}
\end{align}
\end{widetext}

This expression contains traces of products of $4\times 4$ matrices $\hat{v}^x,\hat{G}^{(0)},\hat{v}^y$ and $\hat{G}^{(1)}$.
To complete these traces we decompose the $4\times 4$ matrices into linear combinations of
$\sigma_{\alpha} \tau_{\beta}$,
where $\sigma_\alpha$ and $\tau_\alpha$ are Pauli matrices for the sublattice and particle-hole Nambu subspaces, respectively.
Then
\begin{align}
\hat{v}^x & = v^x_{\alpha} \sigma_\alpha \tau_0 \quad , \quad  \hat{v}^y  = v^y_\alpha \sigma_\alpha \tau_0, \\
\hat{G}^{(0)} & = G^{(0)}_{\alpha \beta} \sigma_\alpha \tau_\beta \; ,  \; \hat{G}^{(1)} = G^{(1)}_{\alpha \beta} \sigma_\alpha \tau_\beta. \label{eq:decomp}
\end{align}
We choose the following basis for the above decomposition
\begin{align}
\sigma_{\alpha}& \equiv(\sigma_0,\sigma_{+},\sigma_{-},\sigma_3), \\
 \tau_{\alpha}& \equiv(\tau_0,\tau_{+},\tau_{-},\tau_3),
\end{align}
where $\sigma_{\pm}=(\sigma_1+i\sigma_2)/\sqrt{2}$ and $\tau_{\pm}=(\tau_1+i\tau_2)/\sqrt{2}$. In Eq.~\eqref{eq:decomp},
and elsewhere, summations over repeated indices  are assumed. In order to extract the coefficients
 $v^x_{\alpha},v^y_{\alpha}, G^{(0)}_{\alpha\beta}$ and $G^{(1)}_{\alpha \beta}$  it will be convenient to introduce both the conjugate of $\sigma_\alpha$,
 denoted as $\bar{\sigma}_\alpha$, and also the conjugate of $\alpha$, denoted as
$\bar{\alpha}$. Their definitions are
\begin{gather}
\bar{\sigma}_\alpha \equiv [\sigma_\alpha]^\dagger =(\sigma_0,\sigma_{-}, \sigma_+,\sigma_3)\equiv \sigma_{\bar{\alpha}}.
\end{gather}
Different components of the $4-$vectors $\sigma_\alpha$ and $\bar{\sigma}_\alpha$ satisfy an orthonormal relation: $\mathrm{Tr}\left\{ \sigma_\alpha \bar{\sigma}_\beta \right\}
= 2 \delta_{\alpha,\beta}$. Using this relation we can obtain the coefficients in Eq.~\eqref{eq:decomp} as follows,
\begin{align}
v^x_\alpha & = \frac{1}{4} \mathrm{Tr}[\hat{v}_x \bar{\sigma}_\alpha \tau_0] \quad , \quad  v^y_\alpha = \frac{1}{4} \mathrm{Tr}[\hat{v}_y \bar{\sigma}_\alpha \tau_0],\\
G_{\alpha\beta}^{(0)} & = \frac{1}{4} \mathrm{Tr}[\hat{G}^{(0)} \bar{\sigma}_\alpha \bar{\tau}_\beta] \; , \;
G_{\alpha\beta}^{(1)}  = \frac{1}{4} \mathrm{Tr}[\hat{G}^{(1)} \bar{\sigma}_\alpha \bar{\tau}_\beta].
\end{align}
Substituting Eq.~\eqref{eq:decomp} into the expression for $\mathcal{F}_{xy}^{(1)}$
in Eq.~\eqref{eq:Fxy1} gives
\begin{widetext}
\begin{align}
\mathcal{F}_{xy}^{(1)} & =
\bigg\{  v^x_\alpha G_{\beta\gamma}^{(0)} v^y_{\alpha^\prime} G^{(1)}_{\beta^\prime \gamma^\prime} \; \mathrm{Tr}\,[\sigma_\alpha \sigma_\beta
\sigma_{\alpha^\prime} \sigma_{\beta^\prime}] \; \mathrm{Tr}\,[\tau_0\tau_\gamma \tau_0 \tau_{\gamma^\prime}]
+ \big\{ (0) \leftrightarrow (1) \big\} \bigg\} -\bigg\{ x \leftrightarrow y \bigg\}, \label{eq:Fxy1-2}
\end{align}
\end{widetext}
where we have suppressed the arguments of the Green's functions. However, it should be kept
in mind that in each of the two-Green's function products, the first Green's function should be evaluated at $(\vec{k}, i\omega_n+i\nu_m)$; while the second
should be evaluated at $(\vec{k}, i\omega_n)$.
The trace over $\tau_{\alpha}$ Pauli matrix products in Eq.~\eqref{eq:Fxy1-2} is trivial: $\mathrm{Tr}\,[\tau_0\tau_\gamma \tau_0 \tau_{\gamma^\prime}]=2\,\delta_{\gamma,\gamma^\prime}$.
The other trace, $\mathrm{Tr}\,[\sigma_\alpha \sigma_\beta \sigma_{\alpha^\prime} \sigma_{\beta^\prime}]$, is nonzero only
for two cases: $(1)$ all four indices, $\{\alpha,\beta,\alpha^\prime,\beta^\prime \}$, are different from each other; $(2)$
the four indices consist of two identical pairs. However, the latter contribution is even with
respect to the interchange $x \leftrightarrow y$ and therefore contributes zero to $\mathcal{F}_{xy}^{(1)}$ after the antisymmetrization,
$-\big\{x\leftrightarrow y\big\}$.
Therefore the only non-zero contribution comes from the case with all four indices different. Because each of the indices,$\{\alpha,\beta,\alpha^\prime,\beta^\prime \}$,
can take four possible values $\{0,+,-,3\}$ there are $4!=24$ different terms in total.
However, half of them are zero because of the following three identities
\begin{subequations}
\begin{align}
G_{+\gamma}^{(0)} G_{-\bar{\gamma}}^{(1)} - G_{-\gamma}^{(0)} G_{+\bar{\gamma}}^{(1)}  +\left\{(0) \leftrightarrow (1) \right\} & =0, \\
G_{-\gamma}^{(0)} G_{3\bar{\gamma}}^{(1)} - G_{3\gamma}^{(0)} G_{-\bar{\gamma}}^{(1)}  + \left\{(0) \leftrightarrow (1) \right\}  & =0, \\
G_{3\gamma}^{(0)} G_{+\bar{\gamma}}^{(1)} - G_{+\gamma}^{(0)} G_{3\bar{\gamma}}^{(1)}  + \left\{(0) \leftrightarrow (1) \right\} & = 0.
\end{align}
\end{subequations}
Then we are left with
\begin{widetext}
\begin{align}
\mathcal{F}_{xy}^{(1)}
& = 8 \bigg\{ \big\{ v^x_{-} v^y_3 -v^x_3 v^y_{-} \big\} \big\{ G_{0\gamma}^{(0)} G_{+\bar{\gamma}}^{(1)} - G_{+\gamma}^{(0)} G_{0\bar{\gamma}}^{(1)} + \big\{(0)\leftrightarrow (1) \big\}\big\}
\nonumber \\
 & \hspace{5mm} + \big\{ v^x_3 v^y_{+} -v^x_{+} v^y_3 \big\} \big\{ G_{0\gamma}^{(0)} G_{-\bar{\gamma}}^{(1)} - G_{-\gamma}^{(0)} G_{0\bar{\gamma}}^{(1)} + \big\{(0)\leftrightarrow (1) \big\} \big\}
 \nonumber \\
 & \hspace{5mm} + \big\{ v^x_{+} v^y_{-} -v^x_{-} v^y_{+} \big\} \big\{ G_{0\gamma}^{(0)} G_{3\bar{\gamma}}^{(1)} - G_{3\gamma}^{(0)} G_{0\bar{\gamma}}^{(1)} + \big\{(0)\leftrightarrow (1) \big\}\big\} \bigg\}.
 \label{eq:Fxy2}
\end{align}
\end{widetext}
In obtaining this equation we have used the trace identity $\mathrm{Tr}[\sigma^0 \sigma^{+} \sigma^{-} \sigma^3]=2$ as well as its permutations.

Next we need to complete the the Matsubara summation $T\sum_{\omega_n}$ in Eq.~\eqref{eq:pxympyx}.
This can be done for each of the three lines in Eq.~\eqref{eq:Fxy2}.
The derivations are quite lengthy, and we do not present them here.
The final results are:
\begin{widetext}
\begin{subequations}
\begin{align}
        T\sum_n  G^{(0)}_{0\gamma} G^{(1)}_{+\bar{\gamma}} - G^{(0)}_{+\gamma} G^{(1)}_{0\bar{\gamma}} +\{ (0) \leftrightarrow (1) \}
 & =     4   i \; \big\{ f_{\vec{k}} d^*_{\vec{k}} - f^*_{\vec{k}} d_{\vec{k}} \big\} \, \xi_{\vec{k}} \, g_{\vec{k}}\, \sqrt{2} \epsilon_{\vec{k}} \,  S_{\vec{k}}(i\nu_m),  \label{eq:G0Gpfinal} \\
     T\sum_n  G^{(0)}_{0\gamma} G^{(1)}_{-\bar{\gamma}} - G^{(0)}_{-\gamma} G^{(1)}_{0\bar{\gamma}} +\{ (0) \leftrightarrow (1) \}
 & =    4  i \; \big\{ f_{\vec{k}} d^*_{\vec{k}} - f^*_{\vec{k}} d_{\vec{k}} \big\} \, \xi_{\vec{k}} \, g_{\vec{k}} \, \sqrt{2} \epsilon^*_{\vec{k}}
 \, S_{\vec{k}}(i\nu_m), \label{eq:G0Gmfinal} \\
      T\sum_n  G^{(0)}_{0\gamma} G^{(1)}_{3\bar{\gamma}} - G^{(0)}_{3\gamma} G^{(1)}_{0\bar{\gamma}} +\{ (0) \leftrightarrow (1) \}
 & =   4 i \; \big\{ f_{\vec{k}} d^*_{\vec{k}} - f^*_{\vec{k}} d_{\vec{k}} \big\} \, \xi_{\vec{k}} \, g_{\vec{k}} \; 2 g_{\vec{k}} \;  S_{\vec{k}}(i\nu_m)
 -   2 i \; \big\{ f_{\vec{k}} d^*_{\vec{k}} - f^*_{\vec{k}} d_{\vec{k}} \big\} \, \xi_{\vec{k}} \, T_{\vec{k}}(i\nu_m) . \label{eq:G0G3final}
\end{align}
\end{subequations}
For brevity we have introduced two frequency dependent functions, $S_{\vec{k}}(i\nu_m)$ and $T_{\vec{k}}(i\nu_m)$, which are defined as
\begin{align}
 S_{\vec{k}}(i\nu_m) & \approx  M_1 - (\xi_{\vec{k}}^2-g_{\vec{k}}^2-|\epsilon_{\vec{k}}|^2) M_2, \label{eq:Sknum}\\
 T_{\vec{k}}(i\nu_m) & \approx \frac{-i\nu_m}{2 E_+ E_- (E_+ + E_-) \left\{ (E_+ + E_-)^2+\nu_m^2 \right\} },  \label{eq:Tknum}
\end{align}
where the $\approx$ sign means only terms of leading order in $f_{\vec{k}}$ and $d_{\vec{k}}$ have been kept.
$M_1$ and $M_2$ are given by
\begin{subequations}
\begin{align}
M_1 &= - i\nu_m \bigg\{  \frac{C_{++}}{4 E_+^2 +\nu_m^2}  +  \frac{C_{--}}{4 E_-^2 +\nu_m^2}  +  \frac{C_{+-}}{(E_+ + E_-)^{2}+ \nu_m^2}
+ \frac{C_{+-}^{\prime}}{\big\{ (E_+ + E_-)^2 + \nu_m^2 \big\}^2} \bigg\}, \label{eq:R32R14-2}  \\
M_2  & =   \frac{- i\nu_m}{E_+ E_-} \bigg\{ \frac{D_{++}}{4 E_+^2 +\nu_m^2}  +  \frac{D_{--}}{4 E_-^2 + \nu_m^2}  +   \frac{D_{+-}}{(E_+ + E_-)^2+\nu_m^2}
+\frac{D_{+-}^{\prime}}{\big\{ (E_+ + E_-)^2 + \nu_m^2 \big\}^2} \bigg\},  \label{eq:R21R03-2}
\end{align}
\end{subequations}
\end{widetext}
where $C_{++}$, $C_{--}$, $C_{+-}$, $C_{+-}^{\prime}$, $D_{++}$, $D_{--}$, $D_{+-}$, and $D_{+-}^{\prime}$ are eight $\nu_m$ independent coefficients.
The expressions for $C_{++}$, $C_{--}$, $C_{+-}$, $ D_{++}$, $D_{--}$, and $D_{+-}$ were given in Eqs.~\eqref{eq:CDcoeff1}-\eqref{eq:CDcoeff5}.
The other two coefficients are as follows
\begin{gather}
C_{+-}^{\prime} = D_{+-}^{\prime}=\frac{-2}{(E_+ + E_-) (E_+ - E_-)^2}.
\end{gather}
Notice that both the $C_{+-}^{\prime}$ term in Eq.~\eqref{eq:R32R14-2}  and the $D_{+-}^{\prime}$ term in Eq.~\eqref{eq:R21R03-2} have a second order pole at
$\nu_m=\pm i(E_+ + E_-)$ on
the complex $\nu_m$ plane; while all other terms have first order poles. The second order poles appear only in the perturbative calculation
but not in the full $\hat{G}$ calculation. Numerically we found that the second order pole contributions to $\sigma_H^{(1)}$
from Eq.~\eqref{eq:R32R14-2} and ~\eqref{eq:R21R03-2}
are negligible at $\omega \gg \alpha$, where $\alpha$ is the SOC coupling strength. Hence we will ignore them hereafter.
Performing a Wick rotation, $i\nu_m \rightarrow \omega + i\delta$, we see that $S_{\vec{k}}(\omega)/\omega$ and $T_{\vec{k}}(\omega)/\omega$
are given by Eqs.~\eqref{eq:Skw} and ~\eqref{eq:Tkw}.

Now inserting the results from Eqs.~\eqref{eq:G0Gpfinal}-\eqref{eq:G0G3final} into the expression for $\mathcal{F}_{xy}^{(1)}$
in Eq.~\eqref{eq:Fxy2} we obtain
\begin{widetext}
\begin{align}
T\sum_n \mathcal{F}_{xy}^{(1)}
& = 64 i \; \big\{ f_{\vec{k}} d^*_{\vec{k}} - f^*_{\vec{k}} d_{\vec{k}} \big\} \, \xi_{\vec{k}} \, g_{\vec{k}} \, S_{\vec{k}}(i\nu_m) \; \vec{h}\cdot\partial_{k_x} \vec{h}\times \partial_{k_y} \vec{h}
+8 \; \big\{ f_{\vec{k}} d^*_{\vec{k}} - f^*_{\vec{k}} d_{\vec{k}} \big\} \, \xi_{\vec{k}} \, T_{\vec{k}}(i\nu_m) \Omega_{xy},
\label{eq:TFxy}
\end{align}
where we have used
\begin{align}
 \vec{h}\cdot\partial_{k_x} \vec{h}\times \partial_{k_y} \vec{h} =  [v_-^x v_3^y - v_3^x v_-^y] \epsilon_{\vec{k}}/\sqrt{2}
+ [v_3^x v_+^y - v_+^x v_3^y] \epsilon^*_{\vec{k}}/\sqrt{2} + [v_+^x v_-^y - v_-^x v_+^y] \, g_{\vec{k}}, \label{eq:tripleh-2}
\end{align}
and also introduced a notation $\Omega_{xy}$ for the following anti-symmetrized velocity factor
\begin{align}
\Omega_{xy} &\equiv - 2 i\, [v_+^x v_-^y - v_-^x v_+^y] = - i[\partial_{k_x} \epsilon_{\vec{k}} \partial_{k_y}\epsilon^*_{\vec{k}} -
\partial_{k_x} \epsilon^*_{\vec{k}} \partial_{k_y}\epsilon_{\vec{k}}].  \label{eq:appOxy}
\end{align}
\end{widetext}
With these compact notations one can substitute $T\sum_n \mathcal{F}_{xy}^{(1)}$ from Eq.~\eqref{eq:TFxy}
back into Eq.~\eqref{eq:pxympyx} and obtain the final expression for the Hall conductivity as a function of frequency given in Eq.~\eqref{eq:sigmaH1-2}.

\section{Asymptotic result for large $\omega$ and sum rules} \label{sec:sumrule}

In this section we compute $ \langle [\hat{J}_x,\hat{J}_y] \rangle$ on the right hand side of Eq.~\eqref{eq:asymptotic1} for the BdG Hamiltonian $\hat{\mathcal{H}}^{(a)}(\vec{k})$ in Eq.~\eqref{eq:Hblock1}
up to first order in $\hat{\mathcal{H}}^\prime$. Denote the basis of the Hamiltonian $\hat{\mathcal{H}}^{(a)}(\vec{k})$ from Eq.~\eqref{eq:Hblock1} as
$\Psi \equiv (\Psi_1,\Psi_2,\Psi_3,\Psi_4)^{T}$. Then the current operator can be written as $\hat{J}_i
= \sum_{\vec{k}}  \sum_{\alpha \beta}  \Psi^\dagger_{\alpha} (\vec{k}) \, v^i_{\alpha \beta}\, \Psi_{\beta}(\vec{k})$, with $i=\{x,y\}$. The velocity
operator matrix $v^i_{\alpha \beta}$ is given in Eq.~\eqref{eq:vxmatrix}.  Using the fact that
the equal time expectation value $\langle \Psi_\alpha^\dagger \Psi_\beta \rangle = T \sum_n \hat{G}_{\beta\alpha}(\vec{k},i\omega_n)$, we obtain
\begin{align}
\langle [\hat{J}_x,\hat{J}_y] \rangle
& = \sum_{\vec{k}} T \sum_n \bigg\{ A \big\{ G_{11} - G_{22} + G_{33} - G_{44} \big\} \nonumber \\
&+ B \big\{ G_{21} + G_{43} \big\}
- B^* \big\{ G_{12} + G_{34} \big\}\bigg\},  \label{eq:JxJyABB}
\end{align}
with $A$ and $B$ given by
\begin{subequations}
\begin{align}
A & = \partial_{k_x} \epsilon_{\vec{k}} \partial_{k_y} \epsilon_{\vec{k}}^* - \partial_{k_x} \epsilon_{\vec{k}}^* \partial_{k_y} \epsilon_{\vec{k}}, \\
B & = 2 (\partial_{k_x} g_{\vec{k}} \partial_{k_y} \epsilon_{\vec{k}} -\partial_{k_x} \epsilon_{\vec{k}} \partial_{k_y} g_{\vec{k}}).
\end{align}
\end{subequations}
On the right hand side of Eq.~\eqref{eq:JxJyABB}
all Green's function matix elements are evaluated at $(\vec{k},i\omega_n)$.

In Eq.~\eqref{eq:JxJyABB} if we use the zeroth order result, $G^{(0)}_{\alpha\beta}$, for  all the Green's function matrix elements
then we obtain $\langle [\hat{J}_x,\hat{J}_y] \rangle^{(0)}=0$. This is consistent with Eq.~\eqref{eq:asymptotic1} and the fact that $\sigma_H^{(0)}(\omega)\equiv 0$.

The nonzero $\langle [\hat{J}_x,\hat{J}_y] \rangle$ comes from the next order contribution: $\langle [\hat{J}_x,\hat{J}_y] \rangle^{(1)}$.
Substituting the matrix elements of the first order Green's function,
$\hat{G}^{(1)} \equiv \hat{G}^{(0)} \hat{\mathcal{H}}^\prime \hat{G}^{(0)}$, into Eq.~\eqref{eq:JxJyABB} and completing the Matsubara summation,
\begin{align}
\langle [J_x,J_y]\rangle^{(1)} &  = i\; \sum_{\vec{k}}  \frac{- 2  i \,\xi_{\vec{k}}  (f_{\vec{k}} d_{\vec{k}}^* -f_{\vec{k}}^* d_{\vec{k}})}{ E_+ E_- (E_+ + E_-)} \nonumber \\
 & \times \bigg\{ \Omega_{xy} + 8 i \, \frac{g_{\vec{k}} \;\vec{h}\cdot \partial_{k_x} \vec{h}\times \partial_{k_y} \vec{h} }{(E_+ + E_-)^2} \bigg\}, \label{eq:JxJy1}
\end{align}
where $\Omega_{xy}$ is defined in Eq.~\eqref{eq:Oxy}.
The remaining $\vec{k}$ summation in Eq.~\eqref{eq:JxJy1} can be evaluated numerically and the final result is $\langle [J_x,J_y]\rangle^{(1)} \approx - i\, 2.2 \times 10^{-5} t^2 \, e^2/(\hbar \, d)$. Then Eq.~\eqref{eq:asymptotic1} becomes
\begin{gather}
\frac{\sigma_H^{(1)}(\omega \rightarrow \infty )}{e^2/\hbar \, d} = \frac{2.2 \times 10^{-5} }{ (\omega/t)^2} +\mathcal{O}(\frac{1}{(\omega/t)^4}). \label{eq:asymptotic2}
\end{gather}

It is also possible to perform the integral in Eq.~\eqref{eq:sumIm} analytically using $\mathrm{Im}\,\sigma_H^{(1)}(\omega)$
from Eq.~\eqref{eq:sigmaH1}. The result is identical to $-i$ times Eq.~\eqref{eq:JxJy1}. Similarly the integral of Eq.~\eqref{eq:sumRe}
can be performed analytically using Eq.~\eqref{eq:sigmaH1-2}-\eqref{eq:F3omega}. The zero result follows from the analytic structure of the $F_i(\vec{k},\omega)$
in Eq.~\eqref{eq:F1omega}-\eqref{eq:F3omega}. We also numerically evaluate the two sides of Eq.~\eqref{eq:sumRe} and \eqref{eq:sumIm}
using the data from Figs.~\ref{fig:ExpIm} and \ref{fig:ExpRe} and confirm that Eqs.~\eqref{eq:sumRe} and \eqref{eq:sumIm} are well satisfied.

\section{Estimation of the NN hopping $t$} \label{sec:testimation}

\begin{figure}[tp]
\centering
\includegraphics[width=0.75\linewidth]{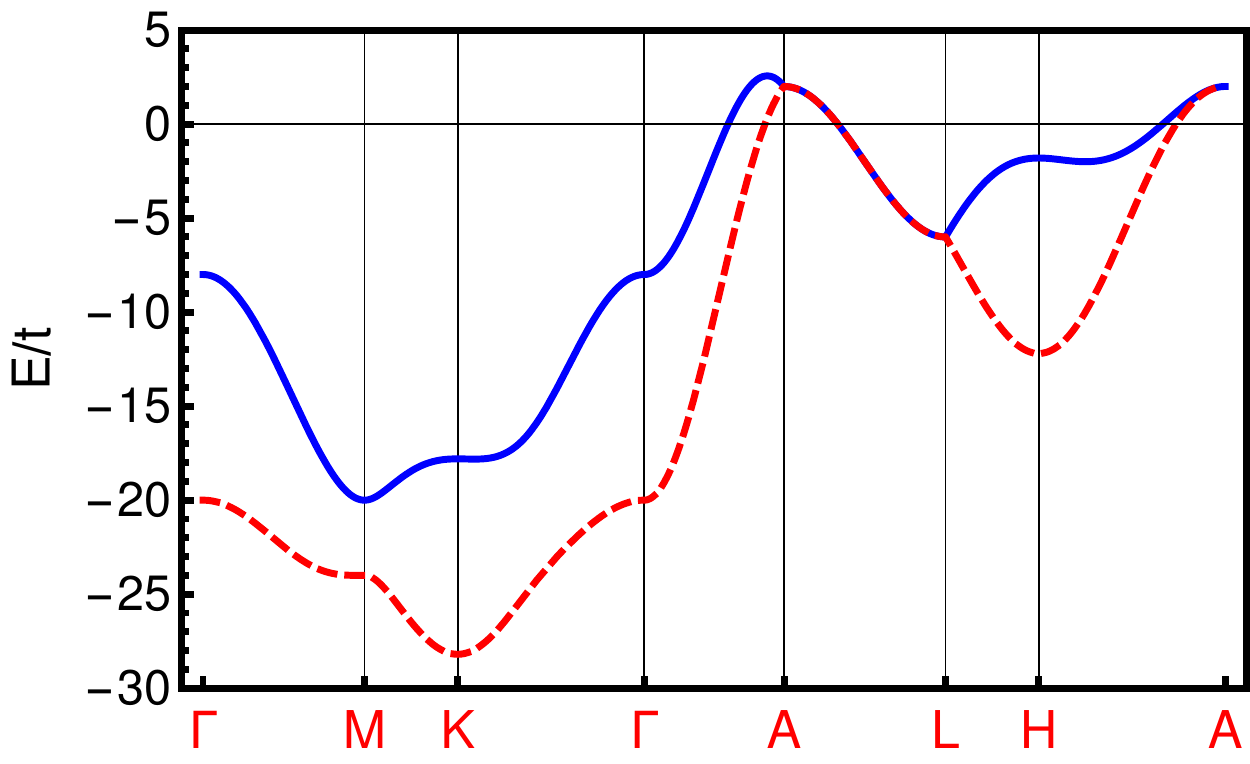}
\caption{Normal state energy dispersions along high symmetry directions of the hexagonal Brillouion zone at $k_z=\pi$. The two energy
band dispersions are $E^{(n)}_{\pm}(\vec{k})=\xi_{\vec{k}}\pm \sqrt{g_{\vec{k}}^2+|\epsilon_{\vec{k}}|^2}$, with $E_{+}^{(n)}$ plotted
in full blue line and $E_{-}^{(n)}$ in the dashed red line. The two bands are degenerate along the symmetry
axis $\mathrm{A-L}$ because $\epsilon_{\vec{k}}=0$ at $k_z=\pi$ and the SOC
vanishes along these directions as well.
}
\label{fig:YanaseGMKGALHA}
\end{figure}

We plot the two normal state energy band dispersions
along high symmetry directions in Fig.~\ref{fig:YanaseGMKGALHA}.
From the dispersions along $\mathrm{A-L-H-A}$, the corresponding band width in the $k_z=0$ plane is
$W \approx 14 t$. We can fit this to the first-principles calculation results from Ref.~\onlinecite{Nomoto2016}.
From the Supplemental Material Fig.S1(b), we estimate that the bandwidth of the dispersions along $\mathrm{A-L-H-A}$ is
$W \approx 0.5 \; \mathrm{eV}$. Therefore, as an estimation,
$ 14 t \approx  0.5 \; \mathrm{eV} \Rightarrow t \approx 36 \; \mathrm{meV}$.

We note that the bands along $\mathrm{\Gamma-M-K-\Gamma}$ in Fig.~\ref{fig:YanaseGMKGALHA} are far below the Fermi energy, which is inconsistent with
the realistic first principle calculation result in Ref.~\onlinecite{Nomoto2016}. This is due to the oversimplification of our model which consists of only
two bands resulting from the $\mathrm{ABAB}$ stacking. Due to this oversimplification, the dispersions along $\mathrm{\Gamma-M-K-\Gamma}$ are not realistic.
In order to estimate how these unrealistic dispersions affect our calculations of $\theta_K$, we have recomputed $\theta_K$
by excluding all $\vec{k}$ points that satisfy $E^{(n)}_{\pm}(\vec{k}) \le E_{-}^{(n)}(\vec{k}=H)$, where $E_{-}^{(n)}(\vec{k}=H)$ is the band bottom of the dispersions along
$\mathrm{A-L-H-A}$ in Fig.~\ref{fig:YanaseGMKGALHA}. The result is similar to the value obatined in the main text without this truncation. In other words, the unrealistic dispersions along $\mathrm{\Gamma-M-K-\Gamma}$ do not significantly change our
conclusion for $\theta_K$. This is because the main contribution to $\sigma_H$ comes from $k_z$ values closer to $k_z=\pi$ and not from the region near $k_z=0$ in the BZ. 


%

\end{document}